\documentclass[superscriptaddress, prd, aps,amsmath,amssymb,showpacs,showkeys, onecolumn]{revtex4-2}
\usepackage[dvips]{graphicx,color}
\usepackage{subfig}
\usepackage{times}
\usepackage{xcolor}
\usepackage[%
  colorlinks=true,
  urlcolor=blue,
  linkcolor=red,
  citecolor=blue
]{hyperref}
\usepackage{orcidlink}

\begin{document}
\title{Geodesics and scalar perturbations of deformed AdS-Schwarzschild black hole with a global monopole surrounded by a quintessence field}

\author{Faizuddin Ahmed\orcidlink{0000-0003-2196-9622}}
\email[Email: ]{faizuddinahmed15@gmail.com}

\affiliation{Department of Physics, Royal Global University, Guwahati, 781035, Assam, India}

\author{Ahmad Al-Badawi\orcidlink{0000-0002-3127-3453}}
\email[Email: ]{ahmadbadawi@ahu.edu.jo}
\affiliation{Department of Physics, Al-Hussein Bin Talal University 71111, Ma’an, Jordan}

\author{ \.{I}zzet Sakall{\i}\orcidlink{0000-0001-7827-9476}}
\email[Email: ]{izzet.sakalli@emu.edu.tr}

\affiliation{Physics Department, Eastern Mediterranean University, Famagusta 99628, North Cyprus via Mersin 10, Turkey}

\author{ Dhruba Jyoti Gogoi \orcidlink{0000-0002-4776-8506}}
\email[Email: ]{moloydhruba@yahoo.in}
\affiliation{Department of Physics, Moran College, Moranhat, Charaideo 785670, Assam, India.}
\affiliation{Research Center of Astrophysics and Cosmology, Khazar University, Baku, AZ1096, 41 Mehseti Street, Azerbaijan.}
\affiliation{Theoretical Physics Division, Centre for Atmospheric Studies, Dibrugarh University, Dibrugarh
786004, Assam, India.\\
{\centering (Dated \today)}
}

\begin{abstract}
We conduct a comprehensive investigation of a deformed AdS-Schwarzschild black hole with a global monopole surrounded by a quintessence field. Our analysis focuses on the geodesic motion of both null and timelike particles, enabling precise determination of the photon sphere and quantification of forces acting on photon particles. We systematically examine how the deformation parameter, control parameter, global monopole parameter, and quintessence field parameter influence the energy, angular momentum, and angular velocities of test particles in circular orbits within the equatorial plane. Additionally, we derive and analyze the perturbative potentials for scalar and electromagnetic fields, demonstrating how these parameters modify their structural profiles. Through time-domain integration, we explore the evolution of scalar and electromagnetic perturbations, revealing distinct phases of transient behavior and quasinormal ringing. Finally, we investigate the emission rates of this black hole, establishing connections between its shadow radius and thermal characteristics. Our findings reveal that the global monopole parameter, deformation parameter and quintessence field enhance black hole stability by reducing emission rates, while the control parameter, which helps avoid central singularities, accelerates evaporation by increasing emission. 
\end{abstract}


\maketitle
\section{Introduction}\label{isec01}
 
The Anti-de Sitter (AdS) spacetime has become increasingly significant in theoretical physics, particularly through the AdS/CFT correspondence, which establishes a remarkable connection between gravitational theories in AdS spacetime and conformal field theories on its boundary \cite{isz01}. This holographic principle has revolutionized our understanding of quantum gravity and provided powerful tools for analyzing strongly coupled quantum systems. Incorporating the AdS background into black hole (BH) solutions enables us to explore thermodynamic properties of BHs from a holographic perspective, potentially shedding light on quantum aspects of gravity \cite{isz02,isz03}. As is well known, BHs serve as cosmic laboratories where the extreme conditions allow us to test fundamental physics theories. In particular, modified BH geometries have garnered significant attention in recent years as they provide valuable insights into quantum gravity effects, the nature of dark energy, and potential resolutions to long-standing theoretical issues such as central singularities. Among these modified BH solutions, the deformed AdS-Schwarzschild BH with a global monopole (GM) surrounded by quintessence field (QF) represents a particularly rich theoretical framework that combines several important physical concepts.

Global monopoles (GMs), first introduced by Barriola and Vilenkin \cite{isz04}, are topological defects arising from the spontaneous breaking of global symmetry in the early universe. These defects produce a solid angle deficit in the surrounding spacetime, modifying the BH geometry in physically meaningful ways. The parameter $\eta$, representing the energy scale of symmetry breaking, determines the strength of this modification. BHs with GMs have been extensively studied due to their unique thermodynamic properties and their potential cosmological implications \cite{isz05,isz06}. QFs have emerged as prominent candidates for explaining the observed accelerated expansion of our universe \cite{isz07}. These dynamical scalar fields, characterized by a negative equation of state parameter ($w < 0$), provide a sophisticated alternative to Einstein's cosmological constant in addressing the dark energy problem. When incorporated into BH solutions, QFs significantly modify both the near-horizon and asymptotic structure of spacetime, leading to interesting phenomenological consequences \cite{isz08,isz09}.

The concept of deformed BHs represents a further refinement in our theoretical understanding of these extraordinary objects. The deformation parameter $\alpha$ introduces corrections that may arise from quantum gravity effects, preserving the asymptotic AdS behavior while modifying the geometry near the BH. These deformations can address theoretical concerns about central singularities and provide a more physically realistic model for BHs in a quantum gravity framework \cite{isz10,isz11}. Similarly, the control parameter $\beta$ offers an additional mechanism to avoid the central singularity at $r=0$, thereby addressing one of the primary theoretical concerns regarding classical BH solutions \cite{isz12}.

Geodesic analysis stands as one of the most powerful tools for investigating the physical properties of BH spacetimes \cite{isz13,isz14}. By studying the paths of both massless (null) and massive (timelike) test particles, we can extract meaningful information about the BH's gravitational field, including the photon sphere, shadow, deflection of light, and stability of circular orbits. These characteristics connect our mathematical models to astrophysical observations, particularly in light of recent advances in BH imaging \cite{isz15,isz16}. Besides, the study of perturbations in BH spacetimes offers crucial insights into their stability and dynamical response to external disturbances \cite{isz17,isz18}. Scalar perturbations, in particular, provide the simplest framework for analyzing how BHs react to small disturbances, revealing information about quasinormal modes, ringdown phases, and late-time tails. These features are directly connected to gravitational wave signals from BH mergers, making perturbation theory increasingly relevant in the era of gravitational wave astronomy \cite{isz19,isz20}. Electromagnetic perturbations extend this analysis to vector fields, offering additional perspectives on BH stability and spectroscopic properties \cite{isz21}. The perturbative potential for both scalar and electromagnetic fields is influenced by the BH parameters, including the deformation parameter $\alpha$, the control parameter $\beta$, the GM parameter $\eta$, and the QF parameters $(c,w)$. Understanding how these parameters affect the perturbative potential provides valuable insights into BH stability across different theoretical frameworks \cite{isz22,isz23}. The time-domain analysis of perturbations complements frequency-domain approaches by capturing the complete temporal evolution of disturbances in BH spacetimes \cite{isz24}. This approach reveals transient behaviors, quasinormal ringing phases, and late-time power-law tails, providing a more comprehensive understanding of BH responses to external perturbations. The numerical integration methods developed by Gundlach et al. \cite{isz25} have proven particularly effective in this context, allowing for accurate simulations of perturbation dynamics.

Hawking radiation and BH emission rates connect classical General Relativity (GR) to quantum field theory, revealing the profound quantum nature of BHs \cite{isz26,isz27}. The emission rate is intrinsically linked to the BH's shadow and thermodynamic properties, highlighting the deep connections between geometry, thermodynamics, and quantum mechanics in BH physics. Investigating how the various parameters in our model affect the emission rate offers insights into BH evaporation, thermodynamic stability, and potential quantum gravity effects \cite{isz28,isz29}. Recent developments in AdS BH thermodynamics have established an extended phase space approach, where the cosmological constant is treated as a thermodynamic pressure, revealing rich phase structures analogous to conventional thermodynamic systems \cite{isz30,isz31}. This approach has uncovered fascinating phenomena such as reentrant phase transitions, multiple critical points, and van der Waals-like behavior in various BH solutions \cite{isz32,isz33}. Incorporating thermal fluctuations further enriches this picture by accounting for quantum corrections to classical thermodynamics, potentially providing glimpses into quantum gravity effects \cite{isz34,isz35}. Several recent studies on black holes, including detailed analyses of their properties and thermodynamic behavior, have been reported in Refs.~\cite{AB3,AB4,AB5,AB6,ref1,ref2,ref3,ref4}.

Motivated by these developments, our present study aims to provide a comprehensive analysis of the deformed AdS-Schwarzschild BH with GM surrounded by QF. We systematically investigate the geodesic motion of test particles, determining key features such as the photon sphere radius and the forces acting on photon particles. We explore how the model parameters influence the energy, angular momentum, and angular velocities of massive test particles in circular orbits. Furthermore, we examine the perturbative potential for both scalar and electromagnetic perturbations, analyzing the time-domain profiles and stability characteristics. Finally, we investigate the thermodynamic properties and emission rates of the BH, establishing connections between its geometric features and quantum behaviors. Overall, our analysis focuses on understanding how the deformation parameter $\alpha$, the control parameter $\beta$, the GM parameter $\eta$, and the QF parameter $c$ (for a specific state parameter $w$) collectively shape the physical properties of the BH \cite{isz36,isz37,isz38,isz39,isz40}.

The paper is organized as follows: In Section \ref{isec02}, we present the static and spherically symmetric deformed AdS-Schwarzschild BH solution with GM surrounded by QF, and conduct a detailed geodesic analysis for both null and timelike particles. Section \ref{isec03} is devoted to the study of scalar perturbations, where we derive the effective potential and analyze its dependence on various parameters. In Section \ref{isec04}, we extend our analysis to electromagnetic (vectorial) perturbations. Section \ref{isec05} presents the time-domain profiles of scalar perturbations, providing insights into the dynamical evolution of disturbances in our BH spacetime. Section \ref{isec06} investigates the emission rate of the BH, establishing connections between its shadow and thermal properties. Finally, in Section \ref{isec07}, we summarize our findings and discuss their implications for BH physics and quantum gravity.

\section{Static and spherically symmetric deformed AdS-Schwarzschild black hole: Geodesics analysis, scalar perturbations} \label{isec02}

In Ref. \cite{MRK}, a static and spherically symmetric deformed BH with thermal fluctuations was analyzed. More recently, the thermodynamics of this BH solution were studied in detail in Ref. \cite{isz36}. Building upon these previous works, the present study extends the analysis by incorporating phantom GMs into the BH solution as well as QF. This extension involves a comprehensive investigation of the geodesic motion of test particles, scalar perturbations, and the emission rate of the system. Specifically, we introduce a static and spherically symmetric deformed AdS-Schwarzschild BH with phantom GMs surrounded by QF, which is described by the following line element:
\begin{equation}
    ds^2=-\mathcal{F}(r)\,dt^2+\frac{dr^2}{\mathcal{F}(r)}+r^2\,(d\theta^2+\sin^2 \theta\,d\phi^2),\label{bb1}
\end{equation}
where the metric function $\mathcal{F}(r)$ (in the natural units) is 
\begin{equation}
    \mathcal{F}(r)=1-8\,\pi\,\eta^2-\frac{2\,M}{r}+\frac{r^2}{\ell^2_{p}}-\frac{c}{r^{3\,w+1}}+\frac{\alpha\,\beta^2}{3\,r\,(\beta+r)^3}+\frac{\alpha}{(\beta+r)^2}\label{bb2}
\end{equation}
with $\alpha$ (with square of length dimension in the natural units) being the deformation parameter that preserves the asymptotic AdS behavior and introduces modifications consistent with additional gravitational fields, $\ell_p$ is the radius of the curvature related with cosmological constant $\Lambda$ as $\frac{1}{\ell^2_{p}}=-\frac{\Lambda}{3}$, $\beta$ (with length dimension) is a constant called control parameter to avoid the central singularity at $r=0$, $\eta$ being the energy scale of the symmetry-breaking \cite{MBAV}, and ($c,w$) are QF parameters. In the limit where $\alpha=0$, that is, without any deformation,  the metric (\ref{bb1}) reduces a spherically symmetric BH with GM surrounded by QF. Moreover, in the limit where $\alpha=0$ and constant $c=0$, the metric (\ref{bb1}) reduces to a spherically symmetric AdS BH solution with a GM reported in Ref. \cite{MBAV}. Moreover, in the limit $\alpha=0$ and $\eta=0$, the metric (\ref{bb1}) reduces to Kiselev-AdS BH reported in Ref. \cite{VVK}, which has widely been investigated in the literature. Additionally, in the limit $\eta=0$ and $c=0$, one can find deformed Schwarzschild BH solution, which has recently been studied in \cite{isz36}.

Taking ordinary derivative of the above function (\ref{bb2}) w. r. t. $r$ and then multiplying both sides by $r$ results
\begin{equation}
    r\,\mathcal{F}'(r)=\frac{2\,M}{r}+\frac{2\,r^2}{\ell^2_{p}}+\frac{c\,(3\,w+1)}{r^{3\,w+1}}-\left(\frac{\alpha\,\beta^2}{3\,r\,(\beta+r)^3}+\frac{\alpha\,\beta^2}{(\beta+r)^4}\right)-\frac{2\,\alpha\,r}{(\beta+r)^3},\label{bb3}
\end{equation}
Thus, we find the quantity $(2\,\mathcal{F}(r)-r\,\mathcal{F}'(r))$ as follows:
\begin{equation}
    2\,\mathcal{F}(r)-r\,\mathcal{F}'(r)=2\,\left[1-8\,\pi\,\eta^2-\frac{3\,M}{r}-\frac{3\,c\,(w+1)/2}{r^{3\,w+1}}+\frac{\alpha}{(\beta+r)^2}+\frac{\alpha\,\beta^2}{2\,r\,(\beta+r)^3}+\frac{\alpha\,r}{(\beta+r)^3}+\frac{\alpha\,\beta^2}{2\,(\beta+r)^4}\right].\label{bb4}
\end{equation}

\subsection{Geodesics Motions}

Geodesic analysis is one of the important tool to investigate the dynamics of test particles around BH. In this section, we investigate motions of photon light and massive particles in the gravitational field around the deformed BH solution (\ref{bb1}) and analyze the outcomes. Geodesics motions in various singular and regular BHs in general relativity have been studied widely (see, for example, Refs. \cite{IJGMMP,IJGMMP2,NPB1,NPB2,CJPHY,EPJC,PDU1,PDU2,NPB3,NPB4} and related references therein).

The Lagrangian density function in terms of the metric tensor $g_{\mu\nu}$ can be expressed as 
\begin{equation}
    \mathcal{L}=\frac{1}{2}\,g_{\mu\nu}\,\dot{x}^{\mu}\,\dot{x}^{\nu},\label{cc1}
\end{equation}
where the dot represents ordinary derivative w. r. t. an affine parameter $\tau$.

Since the selected deformed BH space-time is static and spherically symmetric in nature, we consider geodesics motions in the equatorial plane, defined by $\theta=\pi/2$ and $\dot{\theta}=0$. Therefore, using the metric (\ref{bb1}), we find the Lagrangian density function as,
\begin{eqnarray}
\mathcal{L}=\frac{1}{2}\,\Big[-\mathcal{F}(r)\,\dot{t}^2+\mathcal{F}(r)^{-1}\,\dot{r}^2+r^2\,\dot{\phi}^2\Big]. \label{cc2}
\end{eqnarray}

We see that the Lagrangian density function $\mathcal{L}=\mathcal{L}(r)$ is independent of $(t, \phi)$ coordinates. Thus, there exist two constants of motions that correspond to these cyclic coordinates and these are given by
\begin{eqnarray}
    &&\dot{t}=\frac{\mathrm{E}}{\mathcal{F}(r)}\quad\quad,\quad\quad \dot{\phi}=\frac{\mathrm{L}}{r^2},\label{cc3}
\end{eqnarray}
where $\mathrm{E}, \mathrm{L}$ respectively are the conserved energy and angular momentum.

With these, Eq. (\ref{cc2}) can be written as
\begin{equation}
    \left(\frac{dr}{d\tau}\right)^2+V_\text{eff}(r)=\mathrm{E}^2,\label{cc4}
\end{equation}
which is one-dimensional equation of motion of particles having energy $\mathrm{E}^2$ and $V_\text{eff}(r)$ is the effective potential given by
\begin{equation}
    V_\text{eff}(r)=\left(-\varepsilon+\frac{\mathrm{L}^2}{r^2}\right)\,\mathcal{F}(r)=\left(-\varepsilon+\frac{\mathrm{L}^2}{r^2}\right)\,\left(1-8\,\pi\,\eta^2-\frac{2\,M}{r}+\frac{r^2}{\ell^2_{p}}-\frac{c}{r^{3\,w+1}}+\frac{\alpha\,\beta^2}{3\,r\,(\beta+r)^3}+\frac{\alpha}{(\beta+r)^2}\right),\label{cc5}
\end{equation}
where $\varepsilon=0$ for null geodesics and $-1$ for time-like.

From the above expression (\ref{cc5}), it is evident that the effective potential for null or time-like geodesics in the given gravitational field of BH is influenced by several factors. These include the deformation parameter $\alpha$, the control parameter $\beta$, the symmetry-breaking energy scale parameter $\eta$, and the QF normalization constant $c$ for a specific state parameter $w$. 

\subsection{Null Geodesics: Photon sphere, force on light-like particle and trajectory equation}

Null geodesics represent paths of massless photon particles in a curved space-time. It helps us to study various properties of a BH. These include photon trajectories, BH shadow, deflection of light, and stability of circular null orbits in the given gravitational field.

For light-like geodesics, $\varepsilon=0$, the effective potential from Eq. (\ref{cc5}) reduces as,
\begin{equation}
    V_\text{eff}(r)=\frac{\mathrm{L}^2}{r^2}\,\mathcal{F}(r)=\frac{\mathrm{L}^2}{r^2}\,\left(1-8\,\pi\,\eta^2-\frac{2\,M}{r}+\frac{r^2}{\ell^2_{p}}-\frac{c}{r^{3\,w+1}}+\frac{\alpha\,\beta^2}{3\,r\,(\beta+r)^3}+\frac{\alpha}{(\beta+r)^2}\right),\label{dd1}
\end{equation}

\begin{figure}[ht!]
    \centering
    \subfloat[$\eta^2=0.1,c=0.01,\beta=2$]{\centering{}\includegraphics[width=0.45\linewidth]{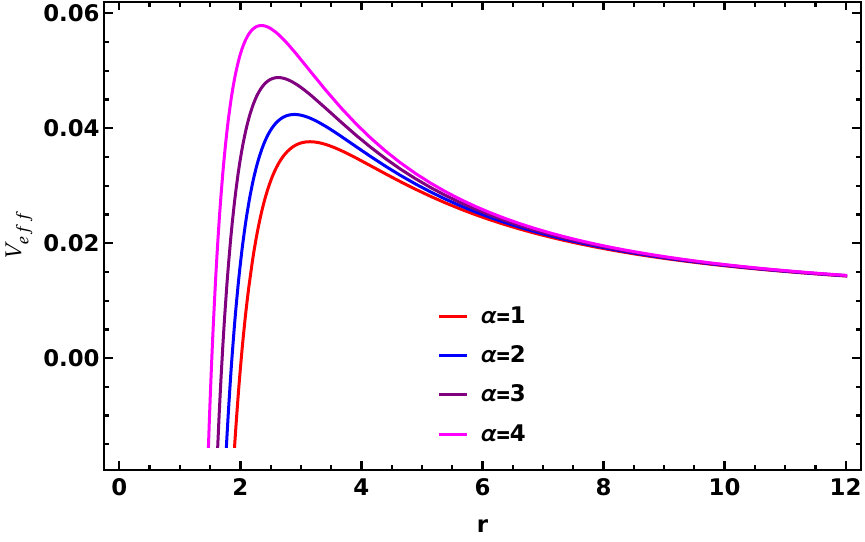}}\quad\quad
    \subfloat[$\eta^2=0.1,c=0.01,\alpha=5$]{\centering{}\includegraphics[width=0.45\linewidth]{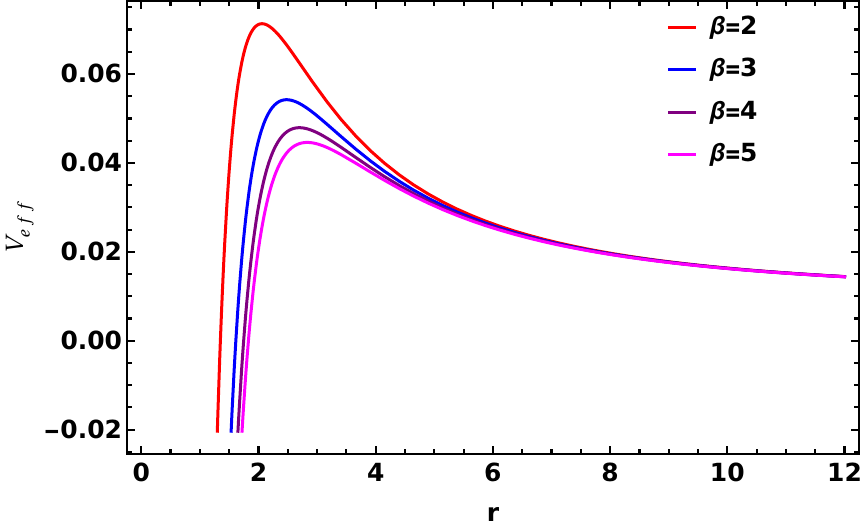}}\\
    \subfloat[$c=0.01,\alpha=5,\beta=2$]{\centering{}\includegraphics[width=0.45\linewidth]{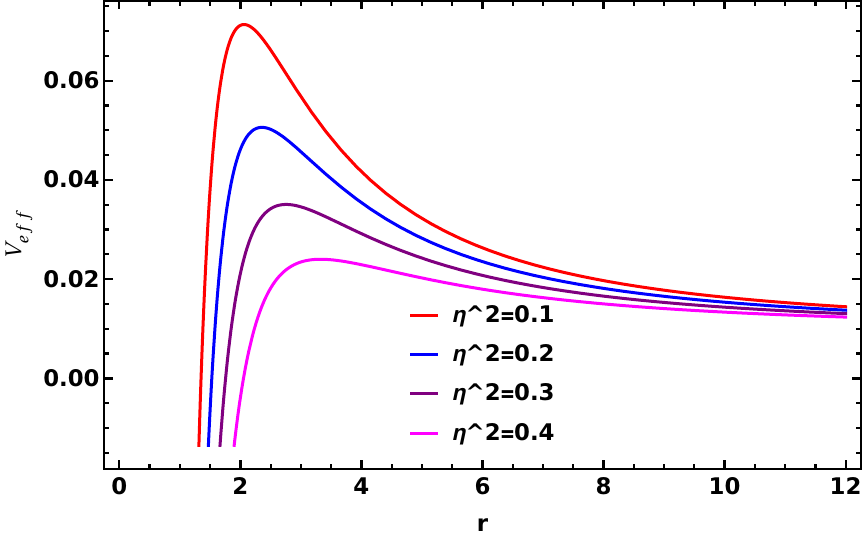}}\quad\quad
    \subfloat[$\eta^2=0.1,\alpha=5,\beta=2$]{\centering{}\includegraphics[width=0.45\linewidth]{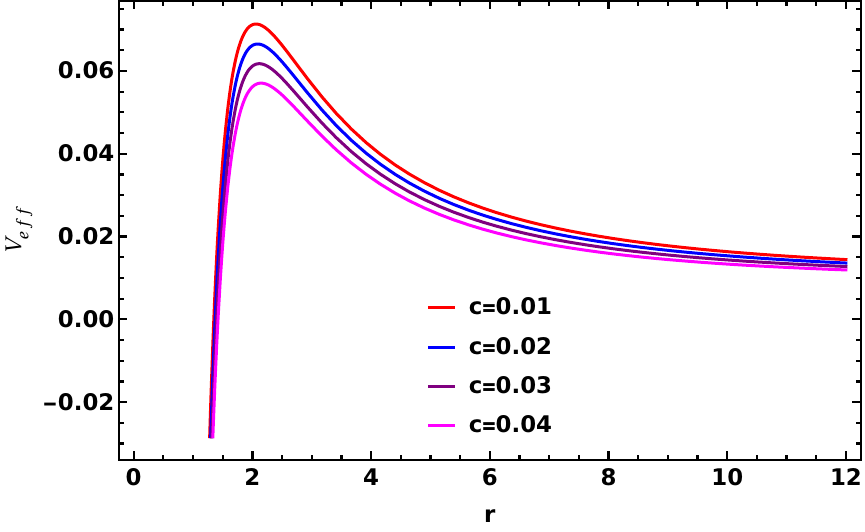}}\\
    \subfloat[$\eta^2=0.1,c=0.01$]{\centering{}\includegraphics[width=0.45\linewidth]{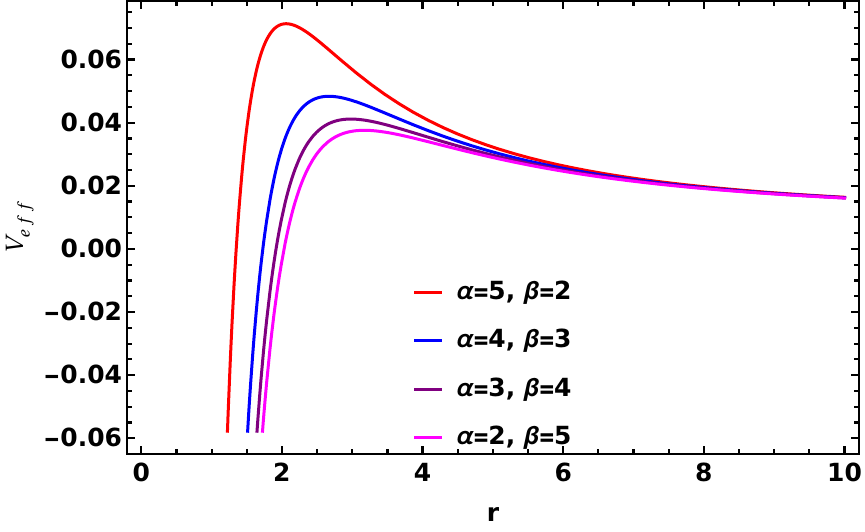}}\quad\quad
    \subfloat[$\alpha=5,\beta=2$]{\centering{}\includegraphics[width=0.45\linewidth]{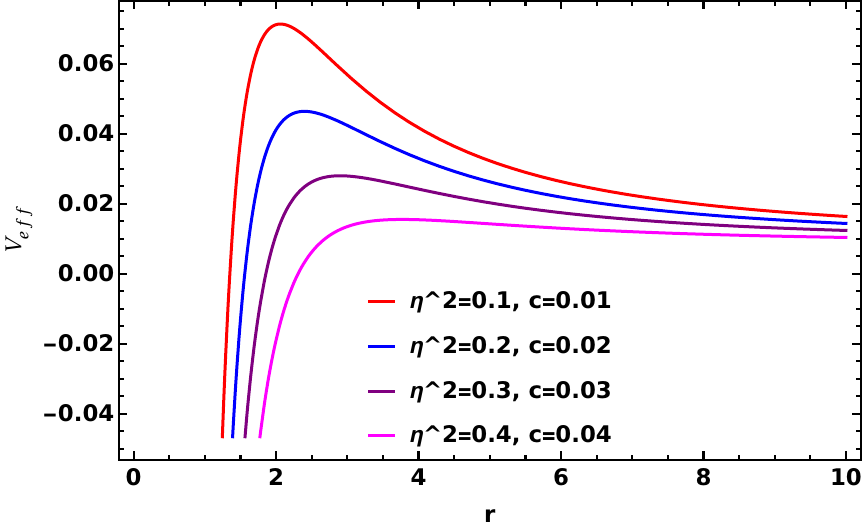}}
    \caption{The behavior of the effective potential for null geodesics. Here, $M=1$, $\mathrm{L}=1$, $\Lambda=-0.03$, and $w=-2/3$.}
    \label{fig:1}
\end{figure}

In Fig. \ref{fig:1}, we present a series of graphs illustrating the behavior of the effective potential for null geodesics under variations of several key parameters. In panel (a), we observe that increasing the deformation parameter $\alpha$ results in a rise in the effective potential. Panel (b) shows a decrease in the potential with increasing values of the control parameter $\beta$. Similarly, in panels (c) and (d), the potential decreases as the symmetry-breaking energy scale $\eta$ and the QF normalization parameter $c$, receptively increase. In panels (e) and (f), the effective potential reduces with the simultaneous increase of both the deformation parameter $\alpha$ and the control parameter $\beta$, as well as with the combined increase of $\eta$ and the QF normalization constant $c$. Overall, across all panels, it is evident that these parameters collectively modulate the effective potential for null geodesics, thereby influencing the deflection of photons in the given gravitational field.

For circular null geodesics, the conditions $\dot{r}=0$ and $\ddot{r}=0$ must hold. These conditions imply the following relations
\begin{equation}
    V_\text{eff}(r)=\mathrm{E}^2\quad\quad \mbox{and} \quad\quad V'_\text{eff}(r)=0,\label{dd2}
\end{equation}
where prime denotes ordinary derivative w. r. to $r$.

The first condition gives us the critical impact parameter for photon light and is given by
\begin{equation}
    \frac{1}{\beta^2_c}=\frac{1}{r_c}\,\left[1-8\,\pi\,\eta^2-\frac{2\,M}{r_c}+\frac{r^2_c}{\ell^2_{p}}-\frac{c}{r^{3\,w+1}_c}+\frac{\alpha\,\beta^2}{3\,r_c\,(\beta+r_c)^3}+\frac{\alpha}{(\beta+r_c)^2}\right].\label{dd3}
\end{equation}

Moreover, the second condition gives us the photon sphere radius $r=r_\text{ph}$ by the following relation
\begin{equation}
    r_\text{ph}\,\mathcal{F}'(r_\text{ph})=2\,\mathcal{F}(r_\text{ph}).\label{dd4}
\end{equation}

The solution to the above equation gives the photon radius depending on the state parameter of the QF. We will focus on the especial case  $w=-2/3$. However, because this equation cannot be solved analytically, it can be calculated numerically. Tables \ref{table1a}, \ref{table2a} and \ref{table3a} show the numerical photon radius values for different deformation parameters $\alpha$, control parameters $\beta$, the energy scale of symmetry-breaking parameter $\eta$ values and the QF normalization parameter $c$.
\begin{center}
\begin{tabular}{|c|c|c|c|c|c|c|c|c|c|c|c|}
 \hline \multicolumn{10}{|c|}{Photon sphere $r_{ph}$ for GM parameter $\eta=0.1$ }
 \\ \hline
  &\multicolumn{3}{|c|}{ $  c=0.01$ } &\multicolumn{3}{|c|}{ $c=0.02$}&\multicolumn{3}{|c|}{   $c=0.03$} \\ \hline 
$\alpha $ & $\beta =0.2$ & $0.4$ & $0.6$  & $0.2$ & $0.4$ & $0.6$ & $0.2$ & $0.4$ & $0.6$ \\ \hline
$0.1$ & $3.0192$ & $3.02718$ & $3.0335$ & $3.06909$ & $3.07713$ & $3.0835$ & 
$3.1225$ & $3.13059$ & $3.13703$ \\ 
$0.3$ & $2.89499$ & $2.92133$ & $2.94167$ & $2.94255$ & $2.96903$ & $2.98955$
& $2.99337$ & $3.02002$ & $3.04075$ \\ 
$0.5$ & $2.76066$ & $2.80955$ & $2.84616$ & $2.80586$ & $2.85497$ & $2.89189$
& $2.85408$ & $2.90345$ & $2.94072$\\
 \hline
\end{tabular}
\captionof{table}{The photon sphere has been tabulated numerically for deformed AdS-Schwarzschild BH with GM surrounded by QF for different values of BH parameters. Setting $\eta=0.1$, $8\,\pi=1$ and $M=1$.} \label{table1a}
\end{center}
\begin{center}
\begin{tabular}{|c|c|c|c|c|c|c|c|c|c|c|c|}
 \hline \multicolumn{10}{|c|}{Photon sphere $r_{ph}$ for GM parameter $\eta=0.3$ }
 \\ \hline
  &\multicolumn{3}{|c|}{ $  c=0.01$ } &\multicolumn{3}{|c|}{ $c=0.02$}&\multicolumn{3}{|c|}{   $c=0.03$} \\ \hline 
$\alpha $ & $\beta =0.2$ & $0.4$ & $0.6$  & $0.2$ & $0.4$ & $0.6$ & $0.2$ & $0.4$ & $0.6$ \\ \hline
$0.1$ & $3.29887$ & $3.3064$ & $3.31247$ & $3.36443$ & $3.37202$ & $3.37815$
& $3.43562$ & $3.44327$ & $3.44948$ \\ 
$0.3$ & $3.17328$ & $3.198$ & $3.21749$ & $3.236$ & $3.26087$ & $3.28056$ & $%
3.30398$ & $3.32904$ & $3.34895$ \\ 
$0.5$ & $3.03823$ & $3.08381$ & $3.11878$ & $3.09808$ & $3.1439$ & $3.17921$
& $3.16283$ & $3.20893$ & $3.24463$\\
 \hline
\end{tabular}
\captionof{table}{The photon sphere has been tabulated numerically for deformed AdS-Schwarzschild BH with GM surrounded by QF for different values of BH parameters. Setting $\eta=0.3$, $8\,\pi=1$ and $M=1$.} \label{table2a}
\end{center}
\begin{center}
\begin{tabular}{|c|c|c|c|c|c|c|c|c|c|c|c|}
 \hline \multicolumn{10}{|c|}{Photon sphere $r_{ph}$ for GM parameter $\eta=0.5$ }
 \\ \hline
  &\multicolumn{3}{|c|}{ $  c=0.01$ } &\multicolumn{3}{|c|}{ $c=0.02$}&\multicolumn{3}{|c|}{   $c=0.03$} \\ \hline 
$\alpha $ & $\beta =0.2$ & $0.4$ & $0.6$  & $0.2$ & $0.4$ & $0.6$ & $0.2$ & $0.4$ & $0.6$ \\ \hline
$0.1$ & $4.05096$ & $4.05751$ & $4.06299$ & $4.17558$ & $4.1822$ & $4.18776$
& $4.31761$ & $4.32432$ & $4.32998$ \\ 
$0.3$ & $3.92215$ & $3.9434$ & $3.96086$ & $4.0422$ & $4.06363$ & $4.08134$
& $4.17866$ & $4.20035$ & $4.21838$ \\ 
$0.5$ & $3.78532$ & $3.82389$ & $3.855$ & $3.90076$ & $3.93963$ & $3.97115$
& $4.03164$ & $4.07091$ & $4.10296$\\
 \hline
\end{tabular}
\captionof{table}{The photon sphere has been tabulated numerically for deformed AdS-Schwarzschild BH with GM surrounded by QF for different values of BH parameters. Setting $\eta=0.5$, $8\,\pi=1$ and $M=1$.} \label{table3a}
\end{center}

\begin{figure}[ht!]
    \centering
    \includegraphics[width=0.65\linewidth]{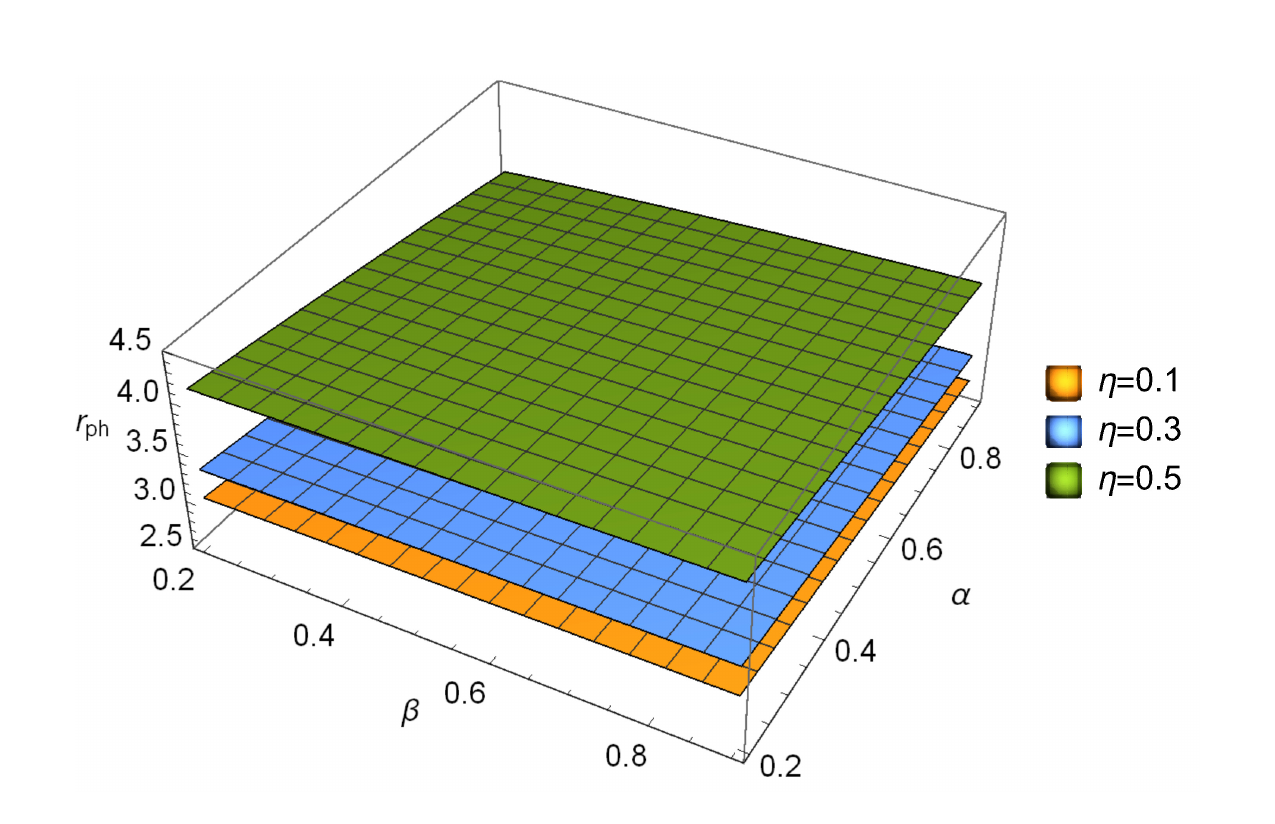}
    \caption{The profile of the photon sphere radius of deformed AdS-Schwarzschild BH with GM surrounded by QF for various values of BH parameters $\alpha$  and $\beta$. $r_{\text{ph}}$ increases with $\eta$. Here, $c=0.02$ and $M=1$.}
    \label{figa1}
\end{figure}

\begin{figure}[ht!]
    \centering
    \includegraphics[width=0.5\linewidth]{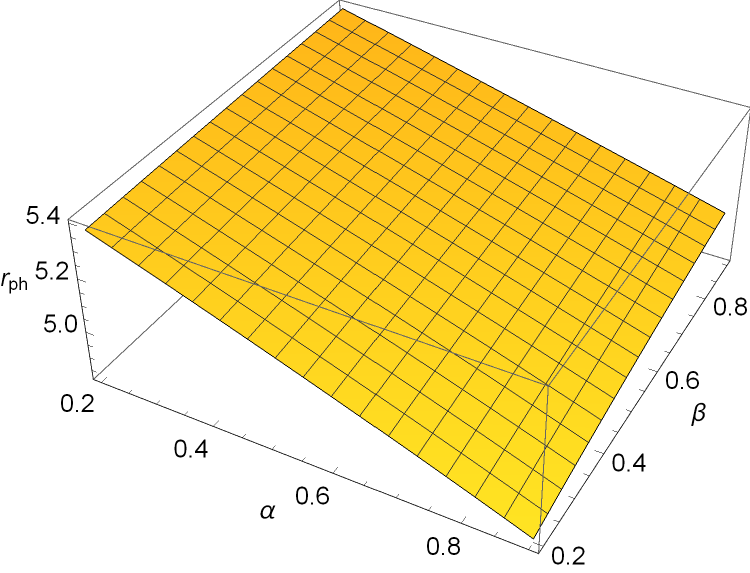}
    \caption{The profile of the photon sphere radius of deformed AdS-Schwarzschild BH with GM surrounded by QF for various values of BH parameters $\alpha$  and $\beta$. $r_\text{ph}$ decreases with  $\alpha$ but increases with $\beta$. Here, $c=0.02$, $\eta=0.4$ and $M=1$.}
    \label{figa2}
\end{figure}

The tables show how the GM parameter $\eta$ affects photon radii. The photon sphere expands as the GM parameter increases.  This implies that the GM parameter has an important role in determining the photon sphere of the AdS-Schwarzschild BH with a GM surrounded by QF. On the other hand, raising the deformation parameter $\alpha$ reduces photon sphere values for fixed control parameters $\beta$. When the deformation parameter is fixed, a similar pattern appears for the control parameter. In Fig. \ref{figa1}, the photon sphere is plotted in three dimensions with respect to the $\beta$  and $\alpha$ parameters for different  GM parameter $\eta$, showing the effect of the $\eta$ parameter and other BH parameters. Fig. \ref{figa2} shows the effect of $\beta$  and $\alpha$ parameters for fixed GM parameter $\eta $. The value of $r_{ph}$ in Fig.s  \ref{figa1} and \ref{figa2} corresponds to the data presented in tables \ref{table1a}, \ref{table2a}, and \ref{table3a}. 

Now, we determine forces on the massless photon light particle when photon light moves under the influence of the gravitational field produced by deformed AdS-Schwarzschild BH with phantom GMs with QF. This force can be obtained using the effective potential $V_\text{eff}$ for light-like geodesics from the following definition:
\begin{equation}
    \mathrm{F}_\text{ph}=-\frac{1}{2}\,\frac{dV_\text{eff}(r)}{dr}.\label{dd5}
\end{equation}

Substituting the effective potential from Eq. (\ref{dd1}) into the relation (\ref{dd5}) results
\begin{equation}
    \mathrm{F}_\text{ph}=\frac{\mathrm{L}^2}{r^3}\,\left[1-8\,\pi\,\eta^2-\frac{3\,M}{r}-\frac{3\,c\,(w+1)/2}{r^{3\,w+1}}+\frac{\alpha}{(\beta+r)^2}+\frac{\alpha\,\beta^2}{2\,r\,(\beta+r)^3}+\frac{\alpha\,r}{(\beta+r)^3}+\frac{\alpha\,\beta^2}{2\,(\beta+r)^4}\right].\label{dd6}
\end{equation}
From the above expression of force (\ref{dd6}), it clearly shows that the force on light-like particle is influenced by the deformation parameter $\alpha$, the control parameter $\beta$, the energy scale of the symmetry breaking parameter $\eta$, and the normalization constant parameter $c$ of QF for a particular state parameter $w$. 

\begin{figure}[ht!]
    \centering
    \subfloat[$\eta^2=0.1,c=0.01,\beta=1$]{\centering{}\includegraphics[width=0.45\linewidth]{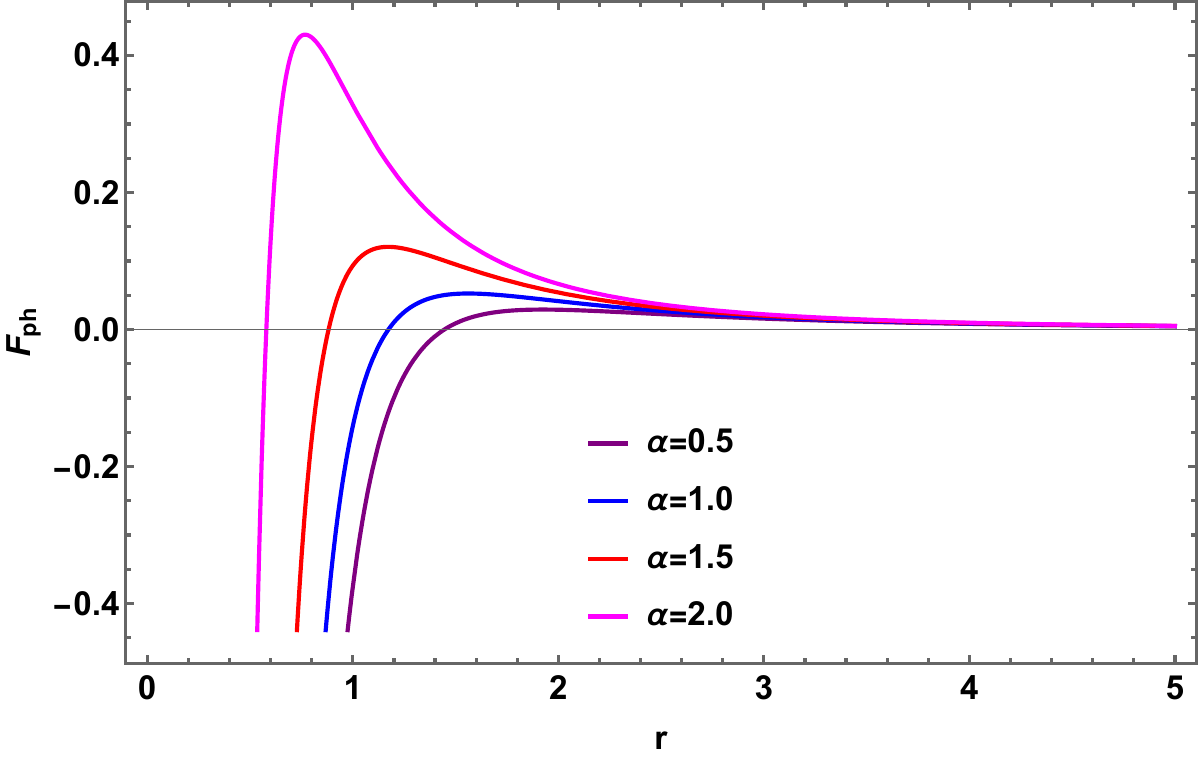}}\quad\quad
    \subfloat[$\eta^2=0.1,c=0.01,\alpha=3$]{\centering{}\includegraphics[width=0.45\linewidth]{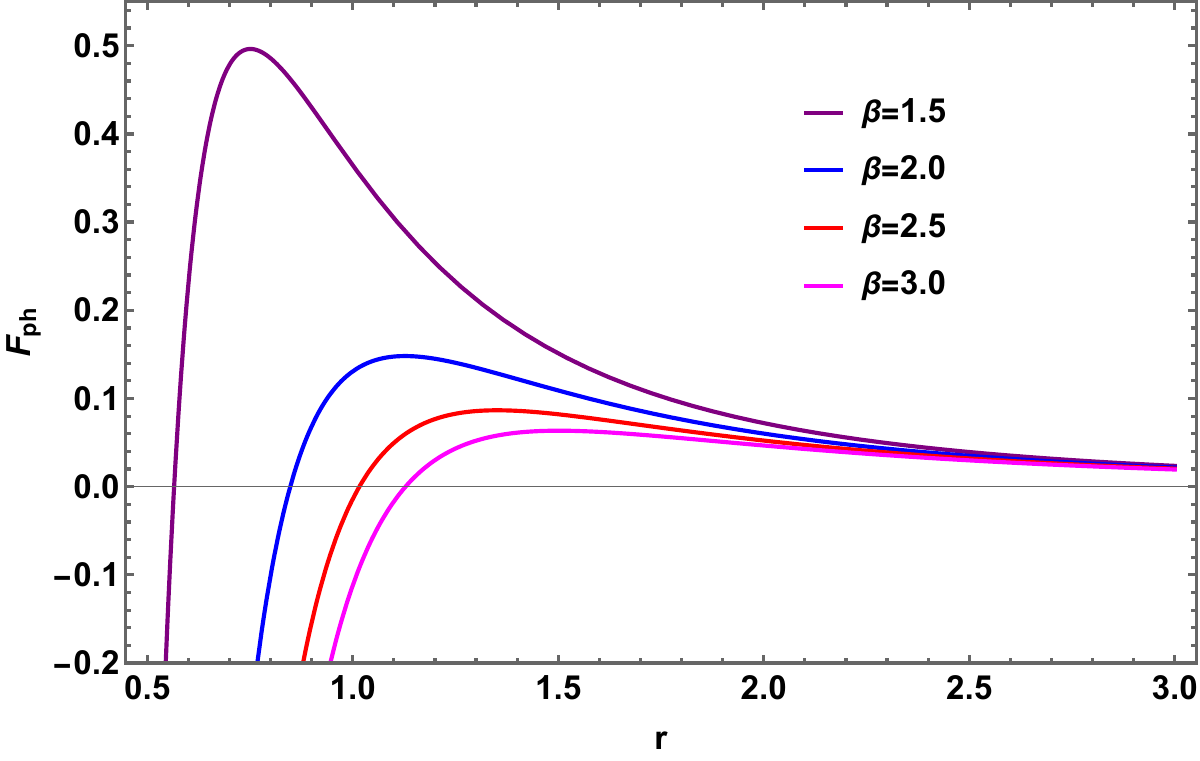}}\\
    \subfloat[$c=0.01,\alpha=5,\beta=2$]{\centering{}\includegraphics[width=0.45\linewidth]{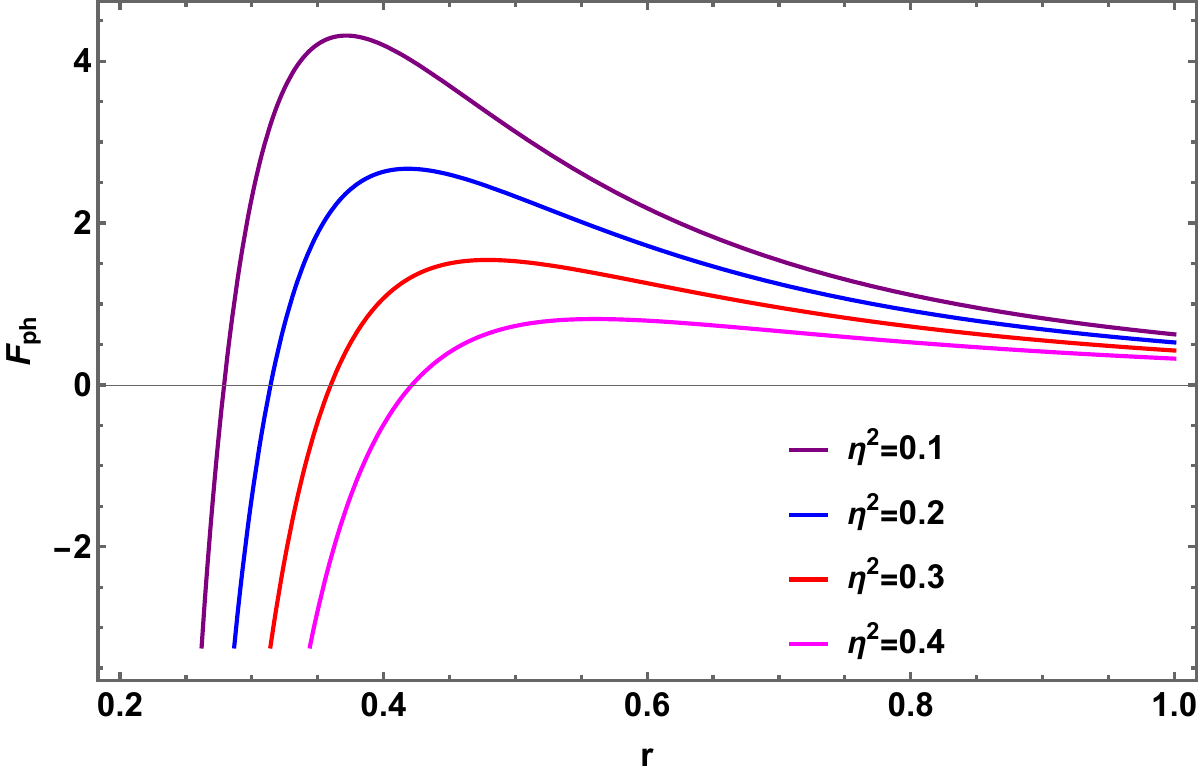}}\quad\quad
    \subfloat[$\eta^2=0.1,\beta=2$]{\centering{}\includegraphics[width=0.45\linewidth]{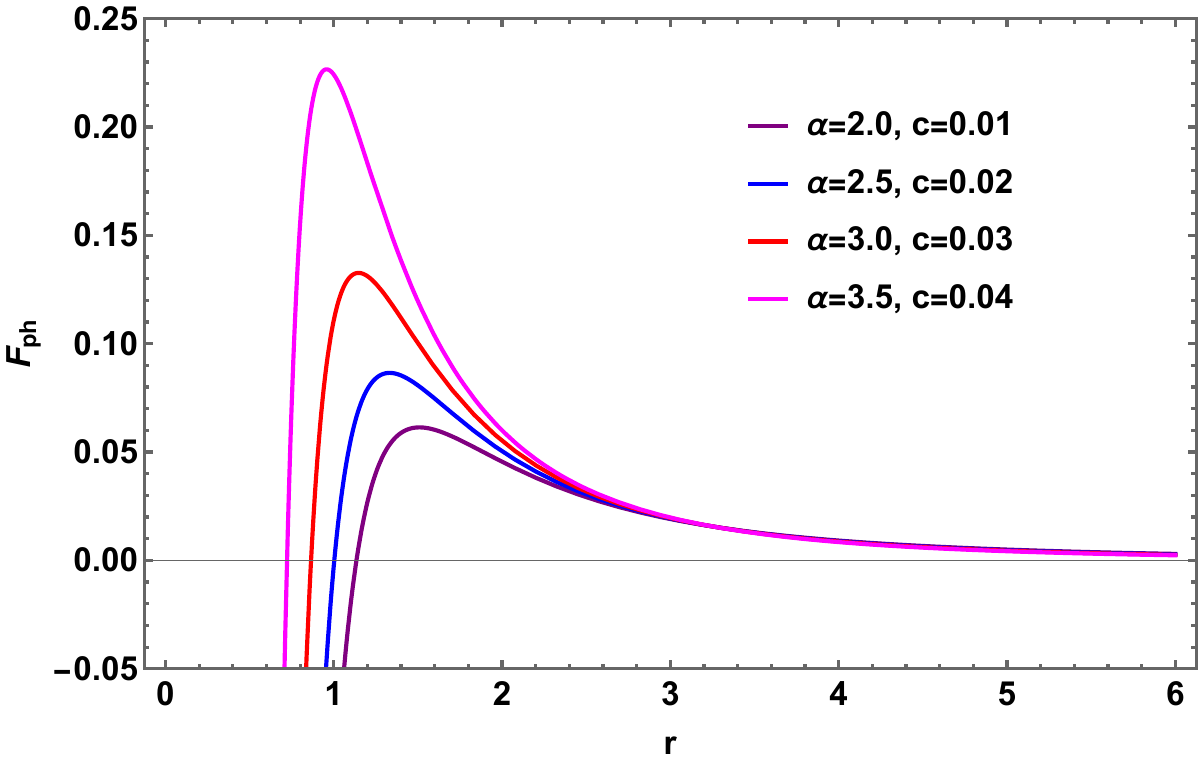}}\\
    \subfloat[$c=0.01$]{\centering{}\includegraphics[width=0.45\linewidth]{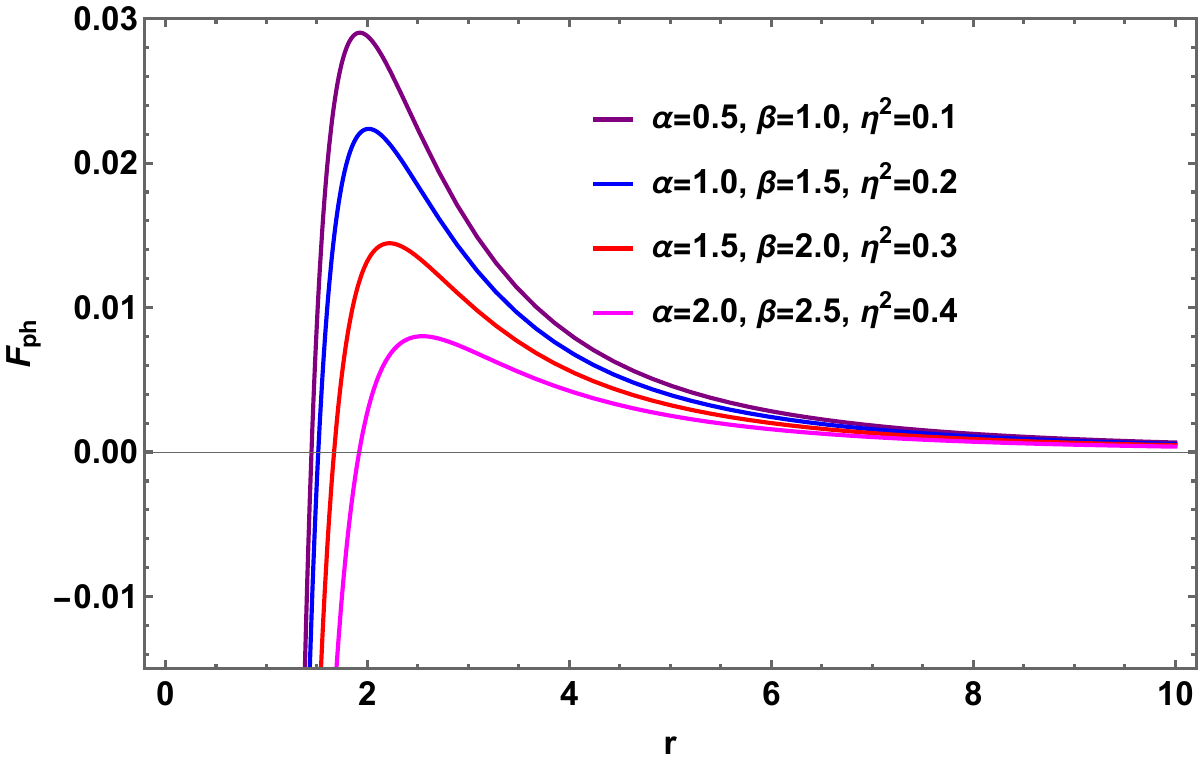}}\quad\quad
    \subfloat[$\alpha=5,\beta=2$]{\centering{}\includegraphics[width=0.45\linewidth]{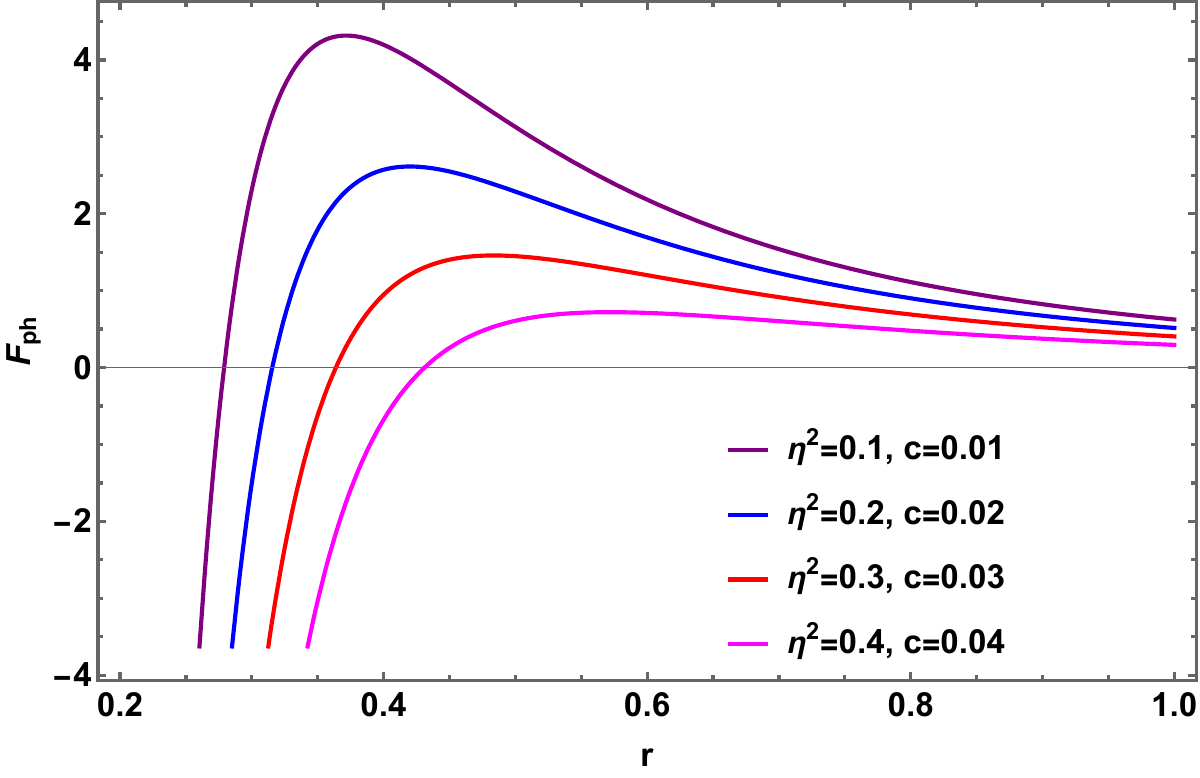}}
    \caption{The behavior of the force on photon particles for different values of $\alpha,\beta,\eta,c$ and their combinations. Here, $M=0.5$, $\mathrm{L}=1$, $\Lambda=-0.03$, and $w=-2/3$.}
    \label{fig:2}
\end{figure}

In Fig. \ref{fig:2}, we present a series of graphs illustrating the behavior of force on photon particles in the given gravitational field under variations of several key parameters. In panel (a), we observe that increasing the deformation parameter $\alpha$ results in a rise in the force. Panel (b) shows a decrease in the force with increasing values of the control parameter $\beta$. Similarly, in panel (c), the force decreases as the symmetry-breaking energy scale $\eta$ increases. Panel (d) shows increase in the force with an increase in the deformation parameter $\alpha$ and the QF normalization parameter $c$. In panels (e) and (f), the force reduces with the simultaneous increase of the deformation parameter $\alpha$, the control parameter $\beta$ and the symmetry-breaking energy scale $\eta$, as well as with the combined increase of $\eta$ and the QF normalization constant $c$, respectively. Noted we have fixed the state parameter $w=-2/3$ for QF, the BH mass $M=0.5$, the cosmological constant $\Lambda=-0.03$, and the angular momentum $\mathrm{L}=1$. Overall, across all panels, it is evident that these parameters collectively alters the dynamics of photon particles, thereby influencing the photon trajectories in the given gravitational field.

\begin{figure}[ht!]
    \centering
    \subfloat[$M=0.1,c=0.01$]{\centering{}\includegraphics[width=0.37\linewidth]{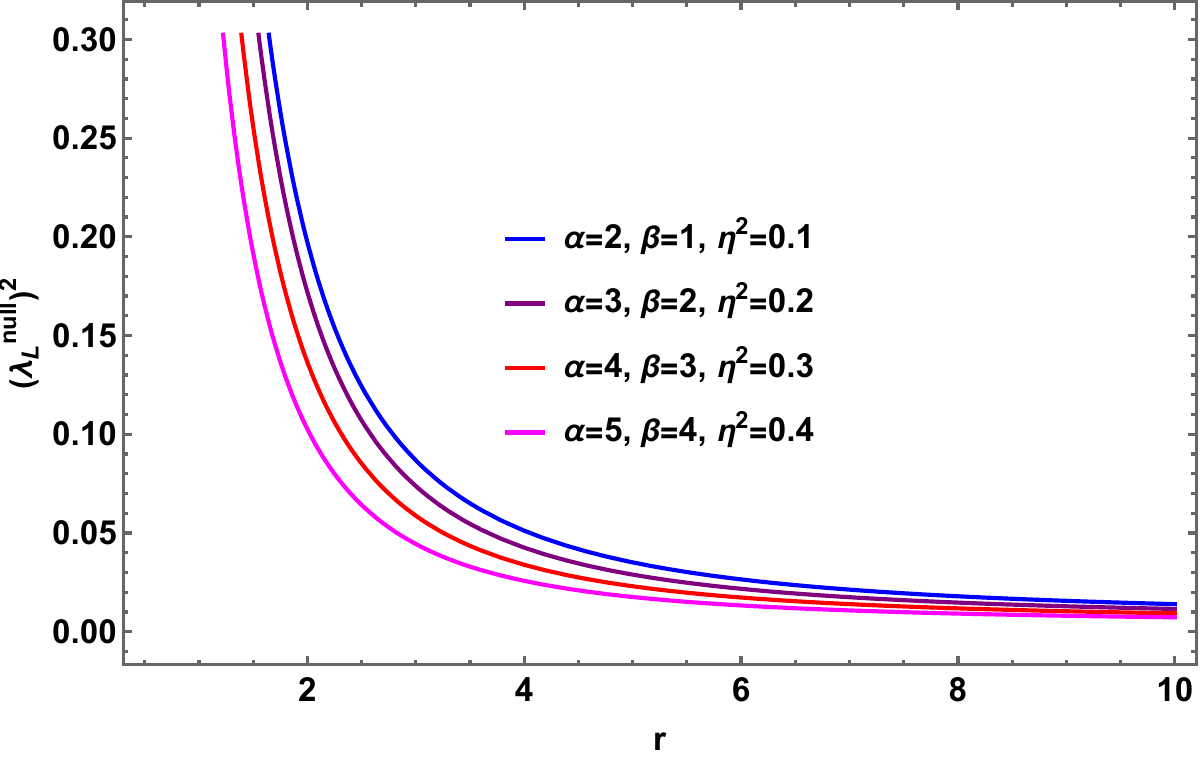}}\quad\quad
    \subfloat[$M=10,c=0.01$]{\centering{}\includegraphics[width=0.37\linewidth]{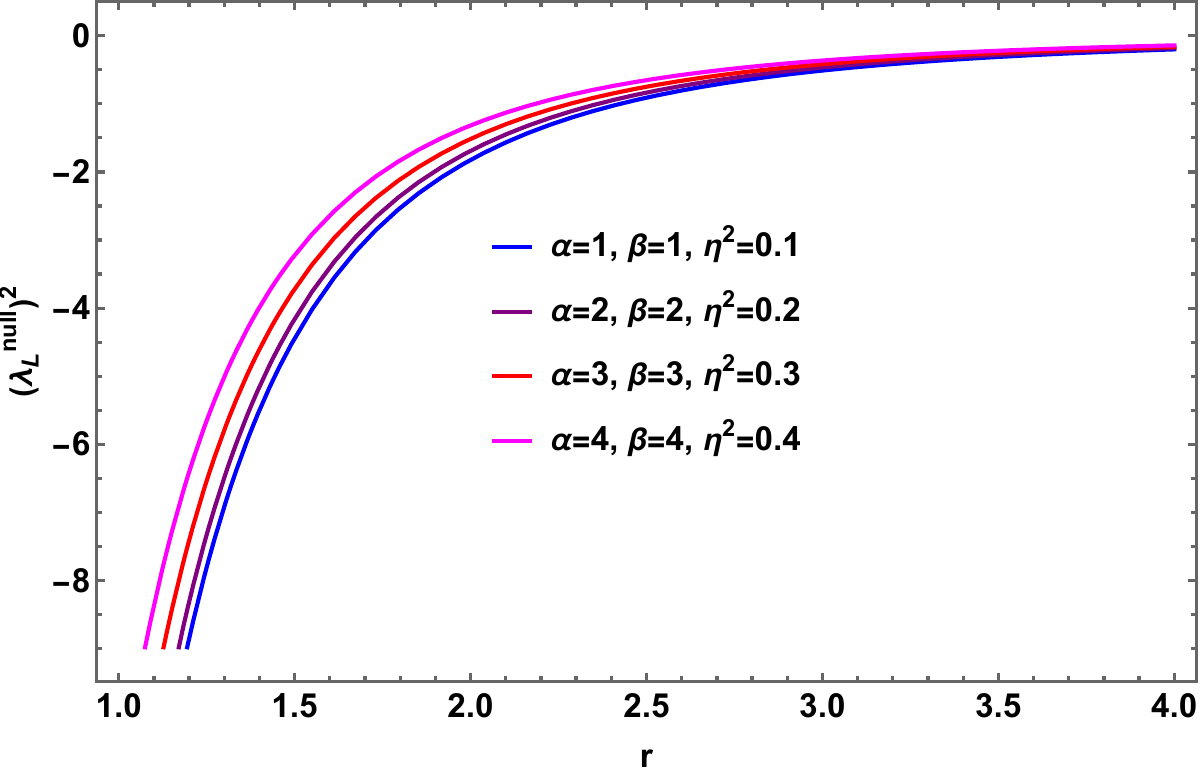}}\\
    \subfloat[$M=0.1,\alpha=5,\beta=2,c=0.01$]{\centering{}\includegraphics[width=0.37\linewidth]{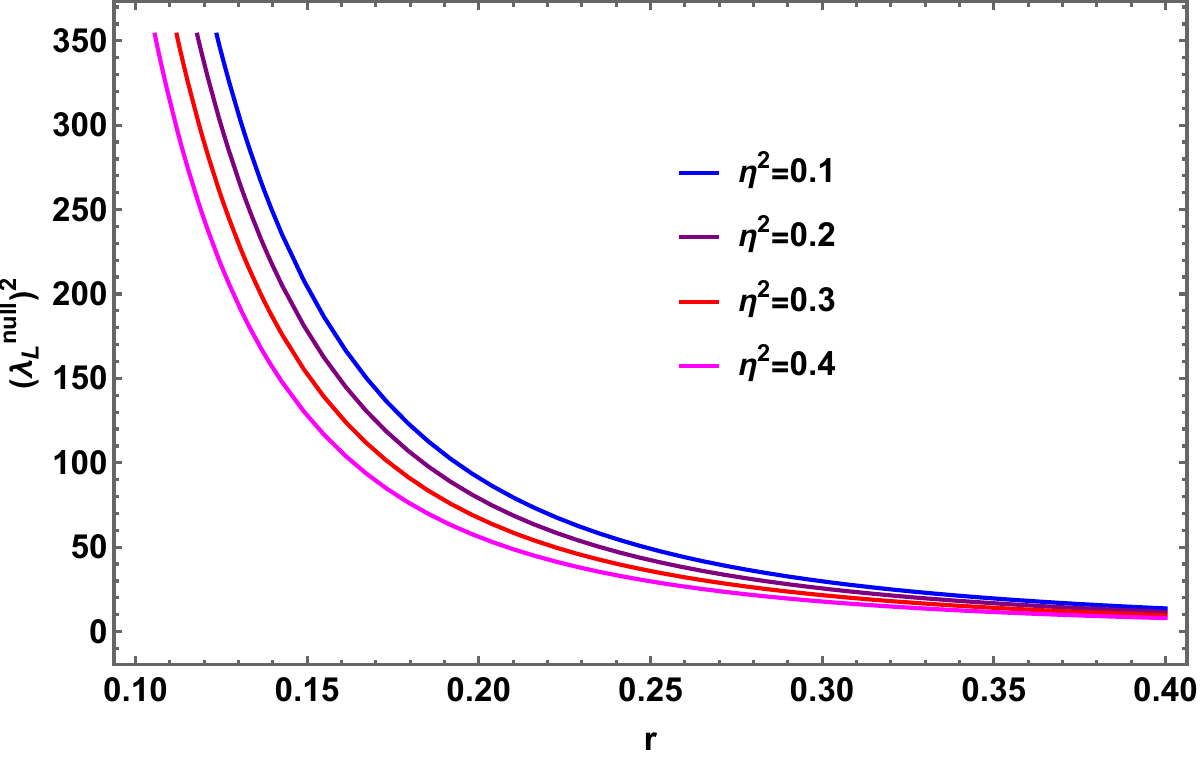}}\quad\quad
    \subfloat[$M=10,\alpha=5,\beta=2,c=0.01$]{\centering{}\includegraphics[width=0.37\linewidth]{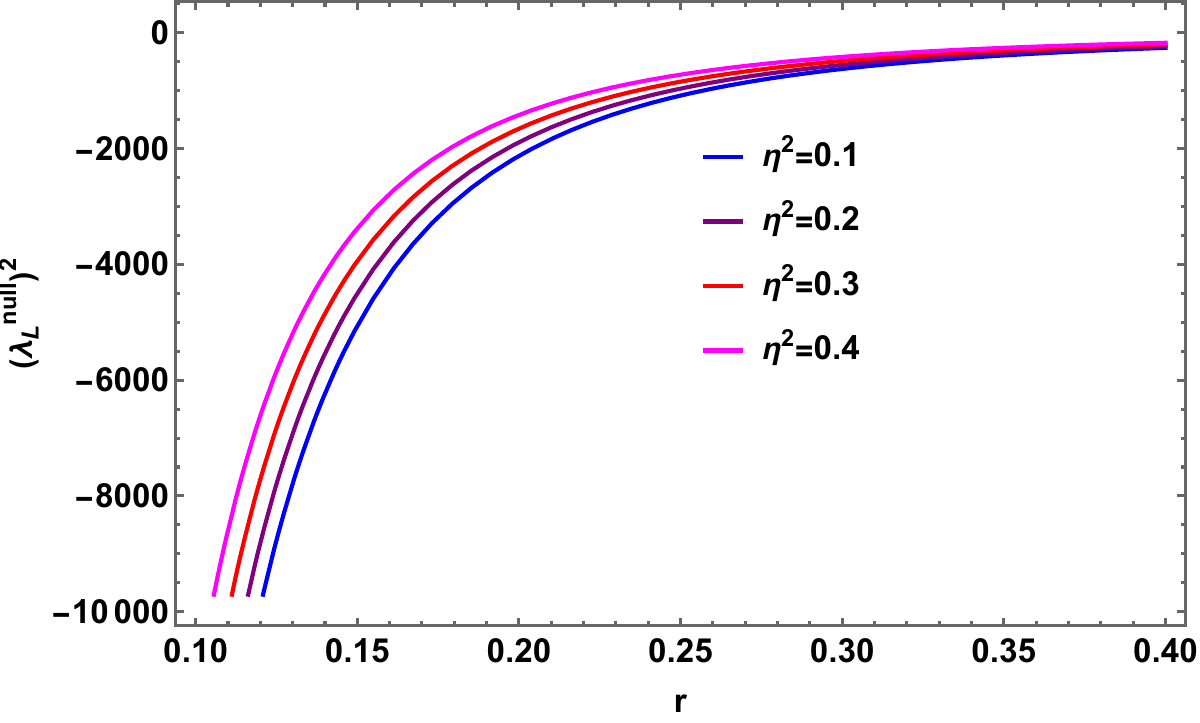}}\\
    \subfloat[$M=0.1,\alpha=5,\beta=2$]{\centering{}\includegraphics[width=0.37\linewidth]{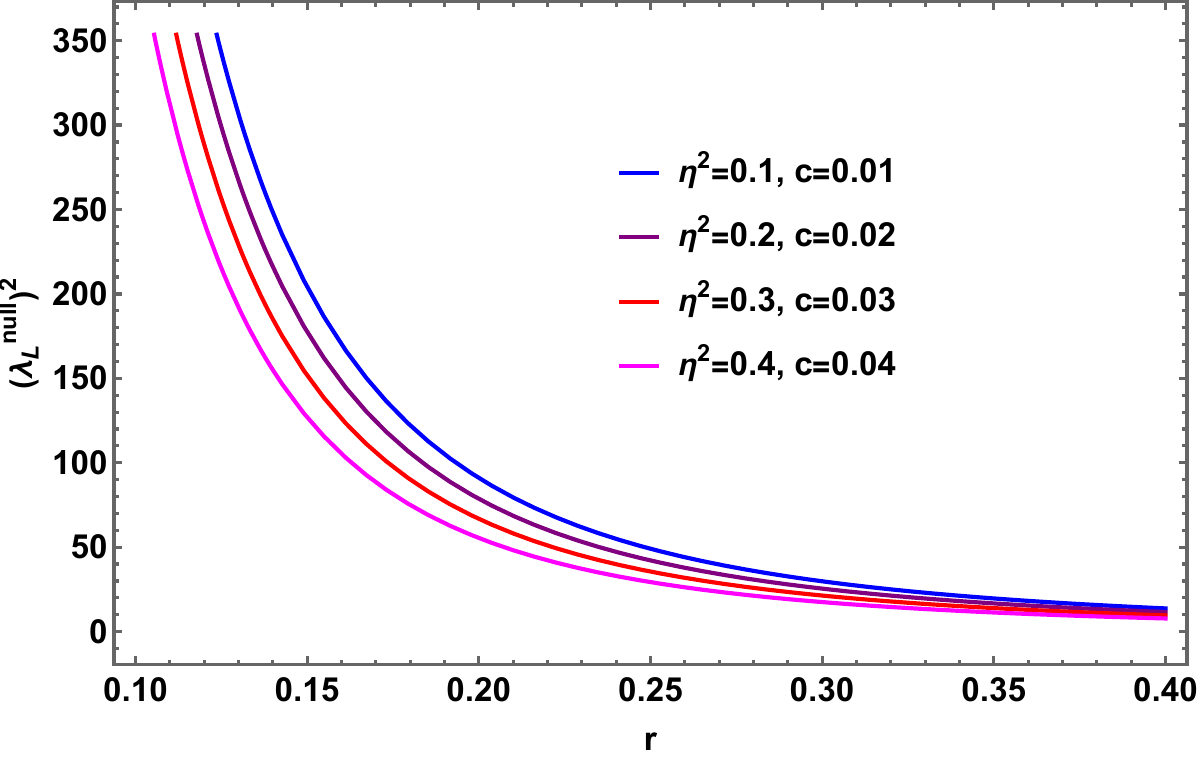}}\quad\quad
    \subfloat[$M=10,\alpha=5,\beta=2$]{\centering{}\includegraphics[width=0.37\linewidth]{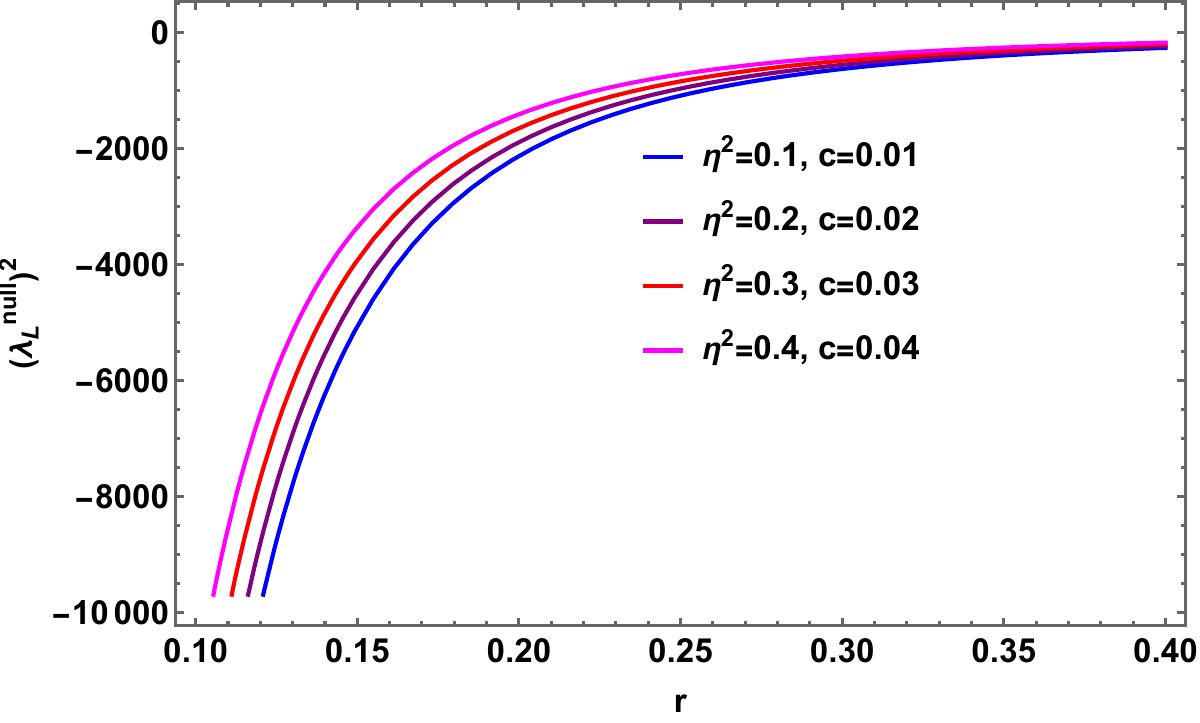}}\\
    \subfloat[$M=0.1,\alpha=5,c=0.01$]{\centering{}\includegraphics[width=0.37\linewidth]{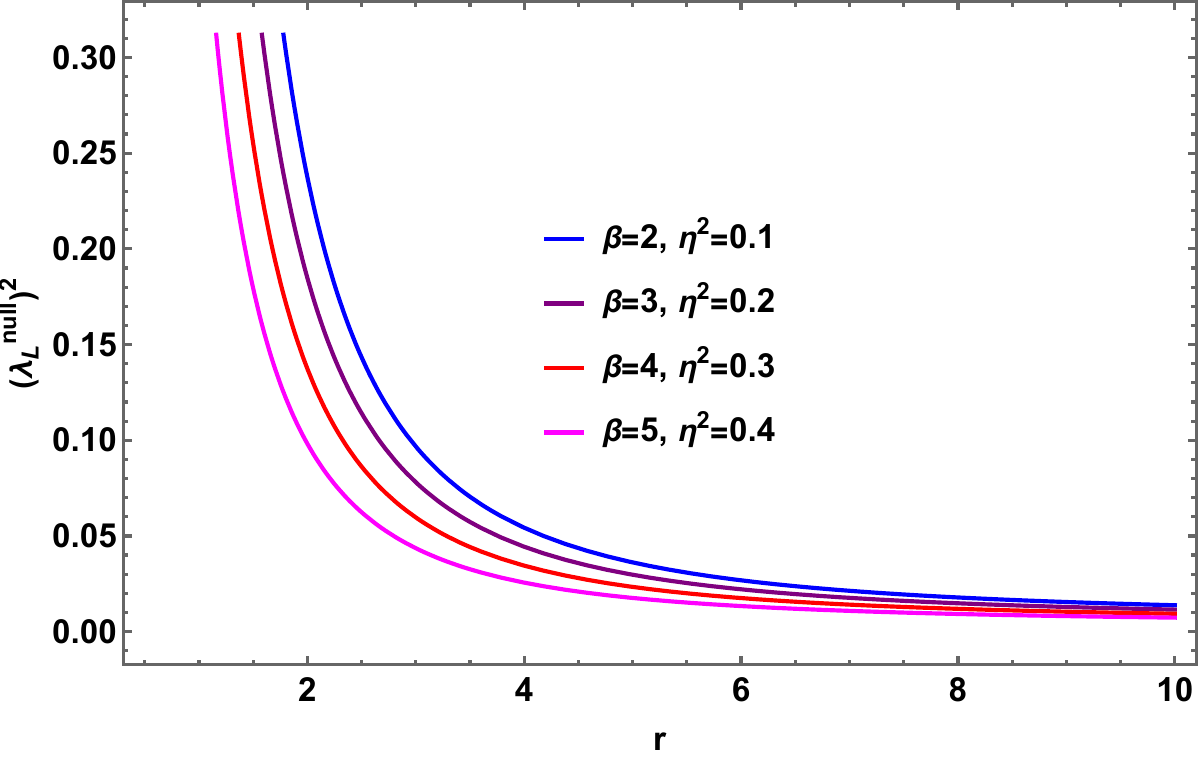}}\quad\quad
    \subfloat[$M=10,\alpha=5,c=0.02$]{\centering{}\includegraphics[width=0.37\linewidth]{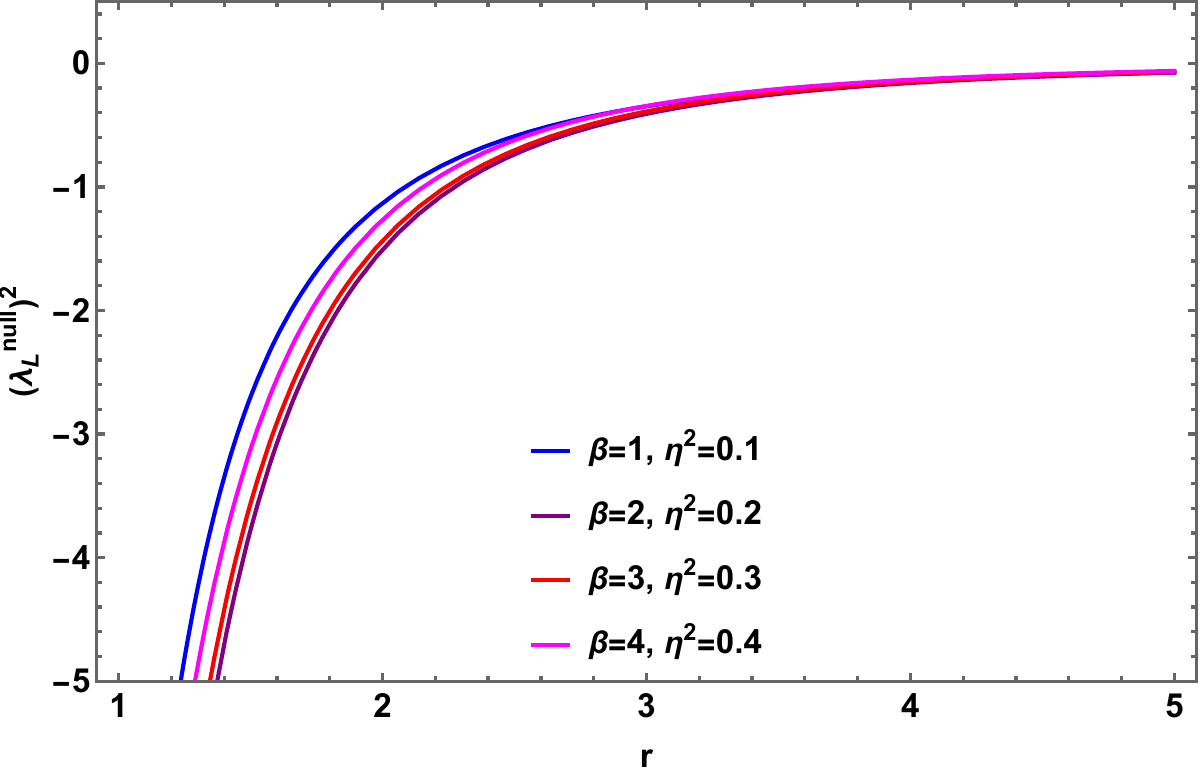}}
    \caption{The behavior of the Lyapunov exponent given in Eq. (\ref{lyapunov4}) for circular null orbits under variations of several key factors in lighter (left panels) and massive (right panels) BH. Here, we set $\Lambda=-0.03$ and $\eta^2=0.01$.}
    \label{fig:3}
\end{figure}

To check stable or unstable circular null geodesics, one can determine the Lyapunov exponent which is defined by \cite{VC}
\begin{equation}
    \lambda^\text{null}_L=\sqrt{-\frac{V''_\text{eff}(r)}{2\,\dot{t}^2}},\label{lyapunov1}
\end{equation}
where $V_\text{eff}(r)$ is given in Eq. (\ref{dd1}) and $\dot{t}$ is given in Eq. (\ref{cc3}).

Using Eqs. (\ref{cc3}) and (\ref{dd4}), we find the following equation of the Lyapunov exponent in terms of the metric function as,
\begin{equation}
    \lambda^\text{null}_L=\sqrt{\mathcal{F}(r)\,\left(\frac{\mathcal{F}(r)}{r^2}-\frac{\mathcal{F}''(r)}{2}\right)}.\label{lyapunov2}
\end{equation}
Substituting the given metric function from Eq. (\ref{bb2}) into the Eq. (\ref{lyapunov2}) and after simplification yields
\begin{eqnarray}
    \lambda^\text{null}_L=\sqrt{1-8\,\pi\,\eta^2-\frac{2\,M}{r}+\frac{r^2}{\ell^2_{p}}-\frac{c}{r^{3\,w+1}}+\frac{\alpha\,\beta^2}{3\,r\,(\beta+r)^3}+\frac{\alpha}{(\beta+r)^2}}\,
    \sqrt{\frac{9\,c\,w\,(w+1)}{2\,r^{3\,(w+1)}}-\frac{2\,r\,\alpha}{(r + \beta)^5} +\frac{1-\eta^2}{r^2}}.\label{lyapunov3}
\end{eqnarray}

From the above expression~(\ref{lyapunov3}), it is evident that the Lyapunov exponent depends on several factors, including the deformation parameter $\alpha$, the control parameter $\beta$, the symmetry-breaking energy scale $\eta$, and the QF normalization constant $c$, for a given state parameter $w$. However, the expression does not make it immediately clear whether the Lyapunov exponent is real and positive or imaginary. It is well known that a positive real Lyapunov exponent indicates unstable circular null orbits, while an imaginary Lyapunov exponent corresponds to stable orbits.

For a specific state parameter, $w=-2/3$, the Lyapunov exponent from Eq. (\ref{lyapunov3}) reduces as
\begin{eqnarray}
    \lambda^\text{null}_L=\sqrt{1-8\,\pi\,\eta^2-\frac{2\,M}{r}+\frac{r^2}{\ell^2_{p}}-c\,r+\frac{\alpha\,\beta^2}{3\,r\,(\beta+r)^3}+\frac{\alpha}{(\beta+r)^2}}\,
    \sqrt{-\frac{c}{r}-\frac{2\,r\,\alpha}{(r + \beta)^5} +\frac{1-\eta^2}{r^2}}.\label{lyapunov4}
\end{eqnarray}

In Fig.~\ref{fig:3}, we present a series of plots that illustrate the behavior of the Lyapunov exponent given in Eq. (\ref{lyapunov4}) associated with circular null orbits, highlighting its dependence on various physical parameters. The panels are organized such that the left column corresponds to a low BH mass, specifically $ M = 0.1 $, while the right column corresponds to a higher BH mass, $ M = 10 $. This arrangement allows for a direct comparison of dynamical stability across different mass regimes. In all left panels, which pertain to the low-mass BH case, the square of the Lyapunov exponent, $ (\lambda^{\text{null}}_L)^2 $, remains strictly positive across variations of the relevant parameters. A positive value of $ (\lambda^{\text{null}}_L)^2 $ implies that small perturbations to the circular null geodesics grow exponentially with time, indicating that these orbits are dynamically unstable. This instability is a well-known feature of null circular orbits around compact objects, particularly in low-mass or near-extremal geometries. In contrast, all right panels, corresponding to the high-mass BH case, exhibit qualitatively different behavior. Here, the square of the Lyapunov exponent becomes negative, meaning $ (\lambda^{\text{null}}_L)^2 < 0 $, which implies that the Lyapunov exponent $ \lambda^{\text{null}}_L $ itself is purely imaginary. This signals the onset of stability in the circular null orbits-perturbations do not grow exponentially but instead, lead to oscillatory behavior. This stable regime is particularly important because it permits the computation of quasinormal mode (QNM) spectra, which characterize the damped oscillations of the spacetime in response to perturbations. These QNMs are essential for understanding the gravitational wave signatures of BHs and can provide direct observational links to the underlying theory of gravity. Overall, the plots in Fig.~(\ref{fig:3}) show that the dynamical stability of circular null geodesics is strongly dependent on the BH mass. Low-mass BHs tend to support unstable photon orbits, while high-mass BHs can stabilize these orbits, enabling a well-defined QNM spectrum. The transition between stability and instability underlines the intricate role of mass and other parameters in the geodesic structure of spacetime.

The coordinate angular velocity in the circular orbits is given by \cite{VC}
\begin{equation}
    \Omega^\text{null}=\frac{\dot{\phi}}{\dot{t}}=\frac{\sqrt{\mathcal{F}(r)}}{r}=\frac{1}{r}\,\sqrt{1-8\,\pi\,\eta^2-\frac{2\,M}{r}+\frac{r^2}{\ell^2_{p}}-\frac{c}{r^{3\,w+1}}+\frac{\alpha\,\beta^2}{3\,r\,(\beta+r)^3}+\frac{\alpha}{(\beta+r)^2}}.\label{lyapunov5}
\end{equation}

From the above expression~(\ref{lyapunov5}), it is evident that the coordinate angular velocity depends on several factors. These include the deformation parameter $\alpha$, the control parameter $\beta$, the symmetry-breaking energy scale $\eta$, the QF normalization constant $c$, and the state parameter $w$ of QF.

Now, we focus on the trajectory of light-like particles in the gravitational field produced by the selected BH. Using Eqs. (\ref{cc3}), (\ref{cc4}) and (\ref{dd1}), we define the following quantity
\begin{equation}
    \frac{\dot{r}^2}{\dot{\phi}^2}=\left(\frac{dr}{d\phi}\right)^2=r^4\,\left[\frac{1}{\gamma^2}-\frac{1}{r^2}\,\left\{1-8\,\pi\,\eta^2-\frac{2\,M}{r}+\frac{r^2}{\ell^2_{p}}-\frac{c}{r^{3\,w+1}}+\frac{\gamma_1}{r\,(\beta+r)^3}+\frac{\gamma_2}{(\beta+r)^2}\right\} \right],\label{dd7}
\end{equation}
where $\gamma=\frac{\mathrm{E}}{\mathrm{L}}$ is the impact parameter for photon light.

Transforming to a new variable via $u=\frac{1}{r}$, the above equation results
\begin{equation}
    \left(\frac{du}{d\phi}\right)^2+(1-8\,\pi\,\eta^2)\,u^2=\frac{1}{\gamma^2}+\frac{1}{\ell^2_{p}}+c\,u^{3\,w+3}+2\,M\,u^3-{\frac{\alpha\,\beta^2\,u^6}{3\,(1+\beta\,u)^3}-\frac{\alpha\,u^4}{(1+\beta\,u)^2}}.\label{dd8}
\end{equation}

Eq.~(\ref{dd8}) describes the trajectory of massless, light-like particles (photons) under the influence of the gravitational field generated by a deformed AdS-Schwarzschild BH with a GM, surrounded by QF. This equation reveals that the path of photon particles is influenced by several key parameters. These include the deformation parameter $ \alpha $, the control parameter $ \beta $, the symmetry-breaking energy scale $ \eta $, and the QF normalization constant $ c $, all considered for a fixed state parameter $ w $. Together, these parameters collectively determine the geometry of null geodesics in this specific spacetime configuration, thereby shaping the behavior of photon trajectories in the presence of modified gravity and quantum corrections. 

In the limit $\beta=0$, corresponding to the absence of the control parameter, the trajectory Eq. (\ref{dd8}) reduces as
\begin{equation}
    \left(\frac{du}{d\phi}\right)^2+(1-8\,\pi\,\eta^2)\,u^2=\frac{1}{\gamma^2}+\frac{1}{\ell^2_{p}}+c\,u^{3\,w+3}+2\,M\,u^3-\alpha\,u^4.\label{dd9}
\end{equation}
Differentiating the above Eq. (\ref{dd9}) w. r. t. $\phi$ and after simplification results the following differential equation
\begin{equation}
    \frac{d^2u}{d\phi^2}+(1-8\,\pi\,\eta^2)\,u=\frac{c}{2}\,(3\,w+3)\,u^{3\,w+2}+3\,M\,u^2-2\,\alpha\,u^3.\label{dd10}
\end{equation}
For a specific state parameter $w=-2/3$, we find the following trajectory equation
\begin{equation}
    \frac{d^2u}{d\phi^2}+(1-8\,\pi\,\eta^2)\,u=\frac{c}{2}+3\,M\,u^2-2\,\alpha\,u^3.\label{dd11}
\end{equation}
The above second-order differential equation is a highly non-linear equation whose exact analytical expression is a challenging task. However, by following approximation scheme, such as $u=u_0+\epsilon\,u_1$, one can derive the angular dependent geodesics path of the radial coordinate $r(\phi)=\frac{1}{u(\phi)}$ which discard here.

Moreover, using Eqs. (\ref{cc3}), (\ref{cc4}) and (\ref{dd1}), we find the non-radial geodesics equation along the time coordinate as follows:
\begin{equation}
    \left(\frac{dr}{dt}\right)^2=\mathcal{F}^2(r)\,\left(1-\frac{\mathcal{F}(r)}{r^2}\,\frac{\mathrm{L^2}}{\mathrm{E}^2}\right).\label{dd12}
\end{equation}
For radial geodesics where the angular momentum vanishes ($\mathrm{L}=0$), we find the following geodesics paths the temporal $(t)$ and radial coordinate $(r)$ as,
\begin{equation}
    \frac{dr}{d\tau}=\pm\,E\quad,\quad \frac{dt}{d\tau}=\frac{\mathrm{E}}{\mathcal{F}(r)}.\label{dd13}
\end{equation}
And the radial motion along the  time coordinate is given by
\begin{equation}
    \frac{dr}{dt}=\pm\,\mathcal{F}(r).\label{dd14}
\end{equation}

\subsection{Time-like Geodesics: dynamics of time-like particles, trajectory equation and Lyapunov exponent}

Timelike geodesics represent the trajectories of massive objects, such as stars and planets, moving in the gravitational field of a BH. The effective potential associated with these timelike geodesics determines the specific energy and angular momentum required for stable orbits around the BH.

Now, we study dynamics of time-like particles in the chosen BH space-time configuration in detail and analyze the result. For time-like geodesics, $\varepsilon=-1$, and hence, the effective potential from Eq. (\ref{cc5}) reduces as 
\begin{equation}
    V_\text{eff}(r)=\left(1+\frac{\mathrm{L}^2}{r^2}\right)\,\mathcal{F}(r),\label{ee1}
\end{equation}

\begin{figure}[ht!]
    \centering
    \subfloat[$\eta^2=0.1,c=0.01,\beta=0.5$]{\centering{}\includegraphics[width=0.45\linewidth]{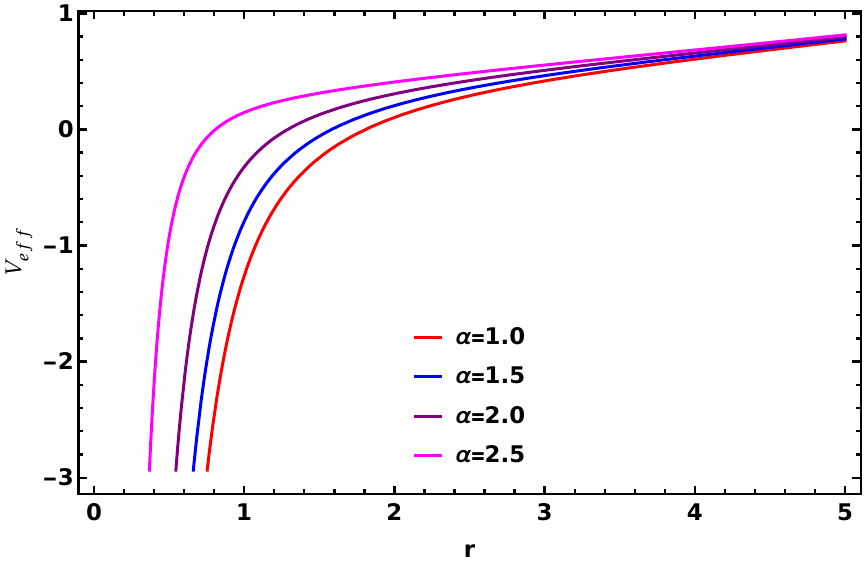}}\quad\quad
    \subfloat[$\eta^2=0.1,c=0.01,\alpha=2$]{\centering{}\includegraphics[width=0.45\linewidth]{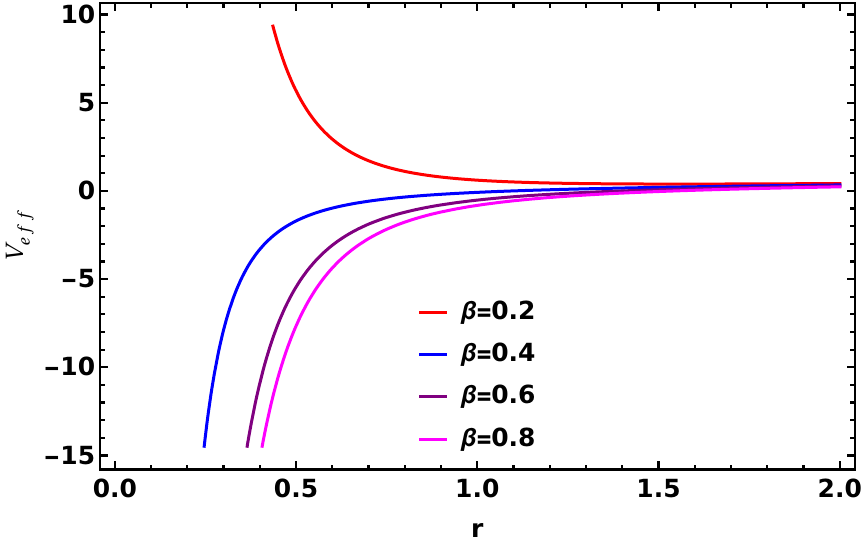}}\\
    \subfloat[$c=0.01,\alpha=5,\beta=2$]{\centering{}\includegraphics[width=0.45\linewidth]{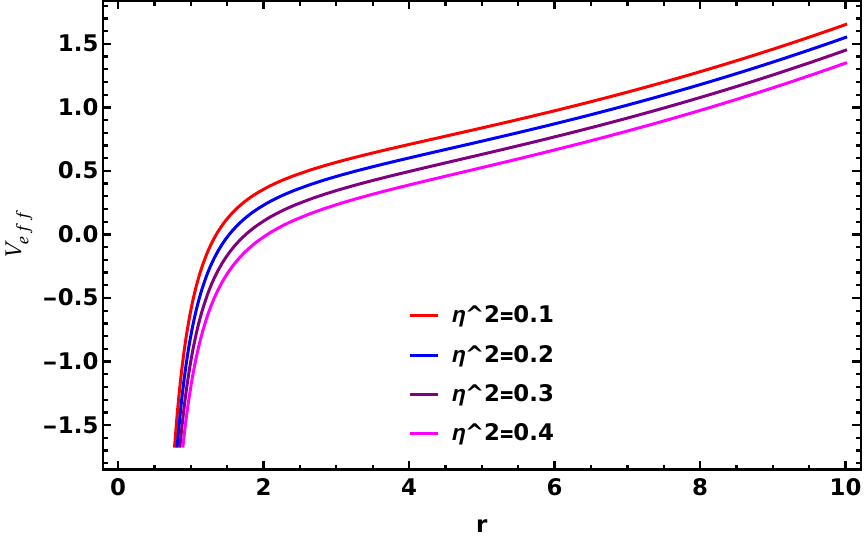}}\quad\quad
    \subfloat[$\eta^2=0.1,\alpha=1,\beta=1$]{\centering{}\includegraphics[width=0.45\linewidth]{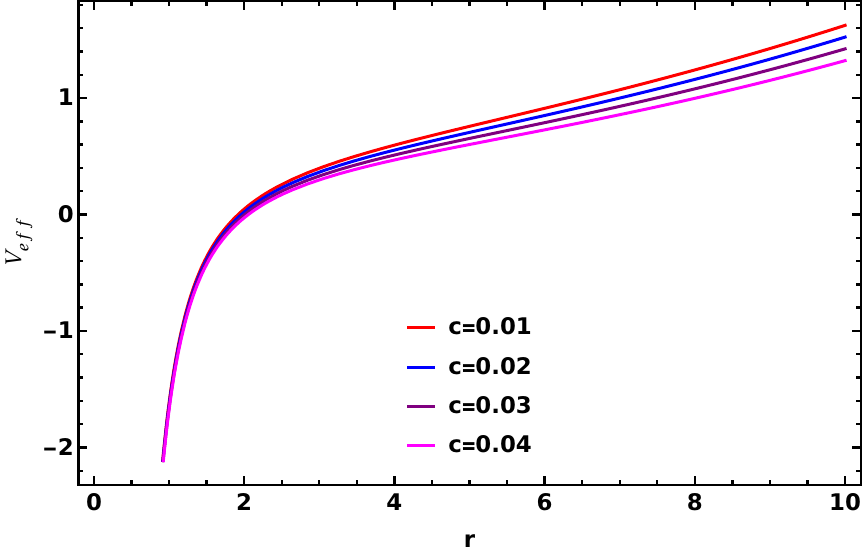}}\\
    \subfloat[$\eta^2=0.1$]{\centering{}\includegraphics[width=0.45\linewidth]{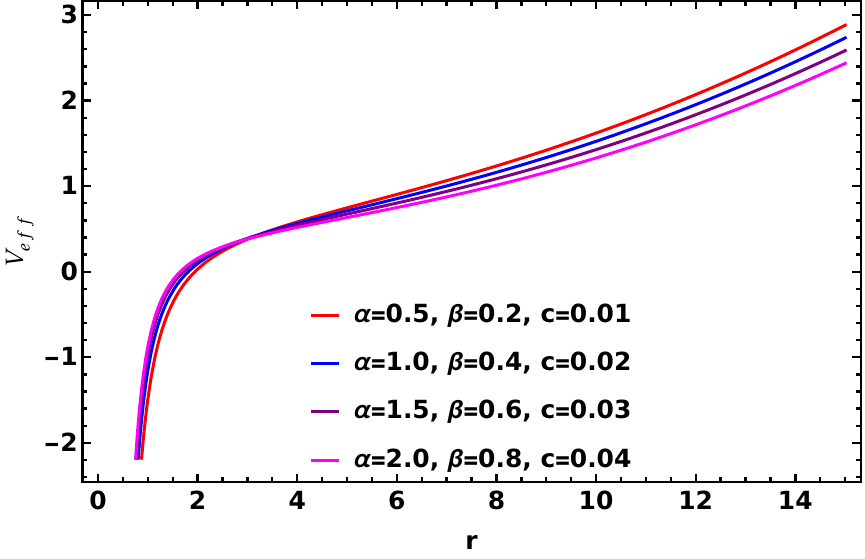}}\quad\quad
    \subfloat[$\alpha=5,\beta=2$]{\centering{}\includegraphics[width=0.45\linewidth]{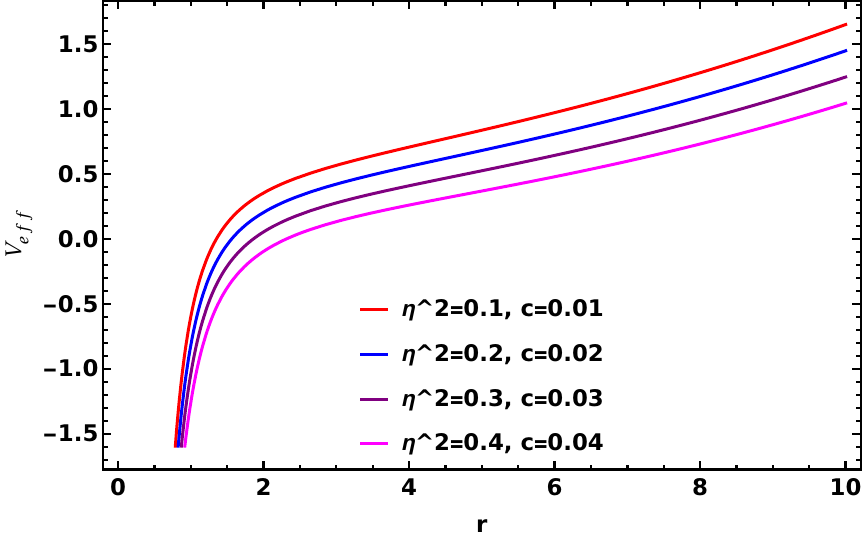}}
    \caption{The behavior of the effective potential for time-like geodesics. Here, $M=1$, $\mathrm{L}=1$, $\Lambda=-0.03$, and $w=-2/3$.}
    \label{fig:4}
\end{figure}

In Fig.~\ref{fig:4}, we present a set of plots depicting the behavior of the effective potential for time-like particles, analyzed under variations of several key parameters. Each panel highlights the influence of a specific parameter or combination of parameters on the potential. In panel (a), we observe that an increase in the deformation parameter $ \alpha $ leads to a rise in the effective potential. Panel (b) demonstrates that the potential decreases with increasing values of the control parameter $ \beta $. Similarly, in panel (c), the effective potential decreases as the symmetry-breaking energy scale $ \eta $ increases, while in panel (d), a similar decline is observed with increasing values of the QF normalization parameter $ c $. Panel (e) illustrates the combined effect of simultaneously increasing $ \alpha $, $ \beta $, and $ c $, which results in an overall decrease in the effective potential. In panel (f), a joint increase in both $ \eta $ and $ c $ also leads to a reduction in the potential. Overall, across all panels, it is evident that these parameters collectively influence the structure of the effective potential for time-like geodesics. Such modifications directly affect the motion of massive particles in this spacetime configuration, altering their orbital characteristics and stability in the given gravitational field.

For circular time-like geodesics, the conditions $\dot{r}=0$ and $\ddot{r}=0$ must holds. These conditions give us the following relations
\begin{equation}
    \mathrm{E}^2=V_\text{eff}(r)=\left(1+\frac{\mathrm{L}^2}{r^2}\right)\,\mathcal{F}(r).\label{ee2}
\end{equation}
And
\begin{equation}
    V'_\text{eff}(r)=0.\label{ee3}
\end{equation}
Moreover, for inner-most stable circular orbits (ISCO) of time-like particles, we must have $V''_\text{eff}\geq 0$.

Substituting the effective potential from Eq. (\ref{ee1}) into the relation (\ref{ee3}) and after simplification results
\begin{eqnarray}
    \mathrm{L}=r\,\sqrt{\frac{\frac{M}{r}+\frac{r^2}{\ell^2_{p}}+\frac{c\,(3\,w+1)/2}{r^{3\,w+1}}-\frac{\alpha\,\beta^2}{6}\,\left(\frac{1}{r\,(\beta+r)^3}+\frac{3}{(\beta+r)^4}\right)-\frac{\alpha\,r}{(\beta+r)^3}}{1-8\,\pi\,\eta^2-\frac{3\,M}{r}-\frac{3\,c\,(w+1)/2}{r^{3\,w+1}}+\frac{\alpha}{(\beta+r)^2}+\frac{\alpha\,\beta^2}{2\,r\,(\beta+r)^3}+\frac{\alpha\,r}{(\beta+r)^3}+\frac{\alpha\,\beta^2}{2\,(\beta+r)^4}}}.\label{ee4}
\end{eqnarray}
Similarly, substituting Eq. (\ref{ee4}) into the Eq. (\ref{ee2}) and after simplification results
\begin{eqnarray}
    \mathrm{E}_{\pm}=\pm\,\frac{1-8\,\pi\,\eta^2-\frac{2\,M}{r}+\frac{r^2}{\ell^2_{p}}-\frac{c}{r^{3\,w+1}}+\frac{\alpha\,\beta^2}{3\,r\,(\beta+r)^3}+\frac{\alpha}{(\beta+r)^2}}{\sqrt{1-8\,\pi\,\eta^2-\frac{3\,M}{r}-\frac{3\,c\,(w+1)/2}{r^{3\,w+1}}+\frac{\alpha}{(\beta+r)^2}+\frac{\alpha\,\beta^2}{2\,r\,(\beta+r)^3}+\frac{\alpha\,r}{(\beta+r)^3}+\frac{\alpha\,\beta^2}{2\,(\beta+r)^4}}}.\label{ee5}
\end{eqnarray}

Eq.s (\ref{ee4}) and (\ref{ee5}) respectively, represents the specific angular momentum and energy of time-like particles orbiting in circular time-like geodesics on the equatorial plane. From the above expressions, one can see that these physical quantities $(\mathrm{L},\mathrm{E})$ are influenced by several factors. These include the deformation parameter $\alpha$, the control parameter $\beta$, the symmetry breaking energy-scale parameter $\eta$, and the QF parameter $(c, w)$.

For a specific state parameter $w=-2/3$, these physical quantities reduce as,
\begin{eqnarray}
    \mathrm{L}&=&r\,\sqrt{\frac{\frac{M}{r}+\frac{r^2}{\ell^2_{p}}-\frac{c\,r}{2}-\frac{\alpha\,\beta^2}{6}\,\left(\frac{1}{r\,(\beta+r)^3}+\frac{3}{(\beta+r)^4}\right)-\frac{\alpha\,r}{(\beta+r)^3}}{1-8\,\pi\,\eta^2-\frac{3\,M}{r}-\frac{c\,r}{2}+\frac{\alpha}{(\beta+r)^2}+\frac{\alpha\,\beta^2}{2\,r\,(\beta+r)^3}+\frac{\alpha\,r}{(\beta+r)^3}+\frac{\alpha\,\beta^2}{2\,(\beta+r)^4}}},\nonumber\\
    \mathrm{E}_{\pm}&=&\pm\,\frac{1-8\,\pi\,\eta^2-\frac{2\,M}{r}+\frac{r^2}{\ell^2_{p}}-c\,r+\frac{\alpha\,\beta^2}{3\,r\,(\beta+r)^3}+\frac{\alpha}{(\beta+r)^2}}{\sqrt{1-8\,\pi\,\eta^2-\frac{3\,M}{r}-\frac{c\,r}{2}+\frac{\alpha}{(\beta+r)^2}+\frac{\alpha\,\beta^2}{2\,r\,(\beta+r)^3}+\frac{\alpha\,r}{(\beta+r)^3}+\frac{\alpha\,\beta^2}{2\,(\beta+r)^4}}}.\label{case}
\end{eqnarray}

Now, we focus into time-like particles trajectory in the given gravitational field produced by the selected BH and show how various parameters influences the trajectory of massive objects. Using Eqs. (\ref{cc3}), (\ref{cc4}) and (\ref{ee1}), we define the following quantity
\begin{equation}
    \frac{\dot{r}^2}{\dot{\phi}^2}=\left(\frac{dr}{d\phi}\right)^2=r^4\,\left[\frac{\mathrm{E}^2}{\mathrm{L}^2}-\left(\frac{1}{\mathrm{L}^2}+\frac{1}{r^2}\right)\,\left\{1-8\,\pi\,\eta^2-\frac{2\,M}{r}-\frac{c}{r^{3\,w+1}}+\frac{r^2}{\ell^2_{p}}+\frac{\gamma_1}{r\,(\beta+r)^3}+\frac{\gamma_2}{(\beta+r)^2}\right\} \right],\label{ee6}
\end{equation}
where $\mathrm{L}$ and $\mathrm{E}$ are given in Eqs. (\ref{ee4}) and (\ref{ee5}), respectively.

Transforming to a new variable via $y=\frac{1}{r}$, the above equation results
\begin{equation}
    \left(\frac{dy}{d\phi}\right)^2=\frac{\mathrm{E}^2(r)}{\mathrm{L}^2(r)}-\left(\frac{1}{\mathrm{L}^2}+y^2\right)\,\left\{1-8\,\pi\,\eta^2-2\,M\,y-c\,y^{3\,w+1}+\frac{1}{y^2\,\ell^2_{p}}+\frac{\alpha\,\beta^2\,y^4}{3\,(1+\beta\,y)^3}+\frac{\alpha\,y^2}{(1+\beta\,y)^2}\right\}.\label{ee7}
\end{equation}
Eq. (\ref{ee7}) represents the trajectory equation for time-like particles under the influence of the gravitational field produced by the deformed AdS-Schwrazschild BH with GM surrounded by QF. From this expression, it is evident that the trajectory of time-like particles is influenced by several factors, including the deformation parameter $\alpha$, the control parameter $\beta$, the symmetry-breaking energy scale parameter $\eta$, and the QF parameters $(c, w)$.

In the limit $\beta=0$, corresponding to the absence of the control parameter, the trajectory Eq. (\ref{ee7}) reduces as
\begin{equation}
    \left(\frac{dy}{d\phi}\right)^2=\frac{\mathrm{E}^2(y)}{\mathrm{L}^2(y)}-\left(\frac{1}{\mathrm{L}^2(y)}+y^2\right)\,\left\{1-8\,\pi\,\eta^2-2\,M\,y-c\,y^{3\,w+1}+\frac{1}{y^2\,\ell^2_{p}}+\alpha\,y^2\right\}.\label{ee8}
\end{equation}
For a specific state parameter $w=-2/3$, we find
\begin{equation}
    \left(\frac{dy}{d\phi}\right)^2=\frac{\mathrm{E}^2(y)}{\mathrm{L}^2(y)}-\left(\frac{1}{\mathrm{L}^2(y)}+y^2\right)\,\left\{1-8\,\pi\,\eta^2-2\,M\,y-c/y+\frac{1}{y^2\,\ell^2_{p}}+\alpha\,y^2\right\}.\label{ee9}
\end{equation}
Substituting $\mathrm{L}$ and $\mathrm{E}$ from Eqs. (\ref{ee4}) and (\ref{ee5}), we get the final trajectory equation for the time-like particles in the given gravitational field.

Let us assume that the time-like particles orbiting around the BH far from the event horizon, $r>>r_\text{h}$. In that case, the gravitational redshift function $-g_{tt}$ can be expressed in terms of the Newtonian gravitational potential $\Phi(r)$ as
\begin{equation}
    \mathcal{F}(r)=1+2\,\Phi(r).\label{ee10}
\end{equation}
Using the metric function given in Eq. (\ref{bb2}), we find this Newtonian gravitational potential $\Phi(r)$
\begin{equation}
    \Phi(r)=\frac{1}{2}\,\left(-8\,\pi\,\eta^2-\frac{2\,M}{r}+\frac{r^2}{\ell^2_{p}}-\frac{c}{r^{3\,w+1}}+\frac{\alpha\,\beta^2}{3\,r\,(\beta+r)^3}+\frac{\alpha}{(\beta+r)^2}\right).\label{ee11}
\end{equation}

Thereby, using this Newtonian potential, one can determine the central force on the particles which is equal to the negative gradient of the potential and is given by
\begin{equation}
    \mathrm{F}_c=-\frac{\partial \Phi(r)}{\partial r}.\label{ee12}
\end{equation}
Using the potential given in Eq. (\ref{ee11}), we find the following expression of the central force
\begin{equation}
    \mathrm{F}_c=-\frac{1}{2}\,\left(\frac{2\,M}{r^2}+\frac{2\,r}{\ell^2_{p}}+\frac{c\,(3\,w+1)}{r^{3\,w+2}}-\frac{\alpha\,\beta^2}{3\,r^2\,(\beta+r)^3}-\frac{\alpha\,\beta^2}{r\,(\beta+r)^4}-\frac{2\,\alpha}{(\beta+r)^3}\right).\label{ee13}
\end{equation}
This central force is balanced by the centripetal force $|\mathrm{F}_c|=\frac{v^2}{r}$, where $v$ is the orbital speed of unit mass particle with which it is orbiting around the BH at far distances. After simplification, we find the orbital speed given by
\begin{equation}
    v=\sqrt{r\,\Big|-\frac{M}{r^2}-\frac{r}{\ell^2_{p}}-\frac{c\,(3\,w+1)}{2\,r^{3\,w+2}}+\frac{\alpha\,\beta^2}{6\,r^2\,(\beta+r)^3}+\frac{\alpha\,\beta^2}{2\,r\,(\beta+r)^4}+\frac{\alpha}{(\beta+r)^3}\Big|}.\label{ee14}
\end{equation}

For a specific state parameter, $w=-2/3$, this orbital speed reduces as
\begin{equation}
    v=\sqrt{r\,\Big|-\frac{M}{r^2}-\frac{r}{\ell^2_{p}}+\frac{c}{2}+\frac{\alpha\,\beta^2}{6\,r^2\,(\beta+r)^3}+\frac{\alpha\,\beta^2}{2\,r\,(\beta+r)^4}+\frac{\alpha}{(\beta+r)^3}\Big|}.\label{ee15}
\end{equation}
From the above expression (\ref{ee15}), it is evident that the orbital speed of time-like particle far from the event horizon of BH is influenced by several key factors which alter the BH geometry. These include the deformation parameter $\alpha$, the control parameter $\beta$, the QF normalization constant $c$, as well as the BH mass $M$ and the radius of curvature $\ell_p$.

\section{Scalar Perturbations} \label{isec03}

Scalar perturbations play a fundamental role in the study of BH spacetimes, particularly in probing their stability under small field fluctuations. This perturbations has been extensively investigated in a wide range of BH solutions within the framework of general relativity, as well as in modified gravity theories. Such studies have yielded significant insights into both the dynamical stability of BH spacetimes and the behavior of scalar fields in curved geometries. Recent research on scalar field perturbations in various BH models can be found in~\cite{NPB2,CJPHY,EPJC,PDU1,PDU2,NPB3,NPB4} and related works.

In this section, we focus on the dynamics of massless scalar field perturbations in a deformed BH solution featuring a GM, surrounded by a QF in AdS background, as given in Eq. (\ref{bb1}). Our analysis begins with the derivation of the massless Klein-Gordon equation, which governs the evolution of scalar fields in the given spacetime geometry. By applying standard techniques of BH perturbation theory, we recast this equation into a Schr\"{o}dinger-like wave equation. We then extract the corresponding effective potential associated with scalar perturbations and analyze how this potential is influenced by several key physical parameters that characterize deviations in the BH geometry. These include the deformation and control parameters, the symmetry-breaking energy scale, and the QF parameters. Through this approach, we aim to understand how scalar field dynamics respond to modifications in the underlying spacetime structure and assess the resulting implications for BH stability.

The massless scalar field wave equation is described by the Klein-Gordon equation as follows Refs. \cite{NPB2,CJPHY,EPJC,PDU1,PDU2}:
\begin{equation}
\frac{1}{\sqrt{-g}}\,\partial_{\mu}\left[\left(\sqrt{-g}\,g^{\mu\nu}\,\partial_{\nu}\right)\,\Psi\right]=0,\label{ff1}    
\end{equation}
where $\Psi$ is the wave function of the scalar field, $g_{\mu\nu}$ is the covariant metric tensor, $g=\det(g_{\mu\nu})$ is the determinant of the metric tensor, $g^{\mu\nu}$ is the contravariant form of the metric tensor, and $\partial_{\mu}$ is the partial derivative with respect to the coordinate systems.

Before, writing explicitly, performing the following coordinate change (called tortoise coordinate) 
\begin{eqnarray}
    dr_*=\frac{dr}{\mathcal{F}(r)}\label{ff2}
\end{eqnarray}
into the line-element Eq. (\ref{bb1}) results
\begin{equation}
    ds^2=\mathcal{F}(r_*)\,\left(-dt^2+dr^2_{*}\right)+\mathcal{D}^2(r_*)\,\left(d\theta^2+\sin^2 \theta\,d\phi^2\right),\label{ff3}
\end{equation}
where $\mathcal{F}(r_*)$ and $\mathcal{D}(r_*)$ are the new metric functions of $r_*$. 

Let us consider the following scalar field wave function ansatz form
\begin{equation}
    \Psi(t, r_{*},\theta, \phi)=\exp(i\,\omega\,t)\,Y^{m}_{\ell} (\theta,\phi)\,\frac{\psi(r_*)}{r_{*}},\label{ff4}
\end{equation}
where $\omega$ is (possibly complex) the temporal frequency, $\psi (r)$ is a propagating scalar field in the candidate space-time, and $Y^{m}_{\ell} (\theta,\phi)$ is the spherical harmonics.

With these, we can write the wave equation (\ref{ff1}) in the following form:
\begin{equation}
    \frac{\partial^2 \psi(r_*)}{\partial r^2_{*}}+\left(\omega^2-\mathcal{V}\right)\,\psi(r_*)=0,\label{ff5}
\end{equation}
where the scalar perturbative potential is given by
\begin{eqnarray}
\mathcal{V}(r)&=&\left(\frac{\ell\,(\ell+1)}{r^2}+\frac{\mathcal{F}'(r)}{r}\right)\,\mathcal{F}(r)\nonumber\\
&=&\left(1-8\,\pi\,\eta^2-\frac{2\,M}{r}-\frac{c}{r^{3\,w+1}}+\frac{r^2}{\ell^2_{p}}+\frac{\alpha\,\beta^2}{3\,r\,(\beta+r)^3}+\frac{\alpha}{(\beta+r)^2}\right)\times\nonumber\\
&&\left\{\frac{\ell\,(\ell+1)}{r^2}+\frac{2\,M}{r^3}+\frac{2}{\ell^2_{p}}+\frac{c\,(3\,w+1)}{r^{3\,(w+1)}}-\frac{\alpha\,\beta^2}{3}\,\left(\frac{1}{r^3\,(\beta+r)^3}+\frac{3}{r^2\,(\beta+r)^4}\right)-\frac{2\,\alpha}{r\,(\beta+r)^3}\right\},\quad \ell\geq 0.\label{ff6}
\end{eqnarray}

Eq. (\ref{ff6}) is the perturbative scalar potential for scalar field in a deformed Schwrazschild AdS BH with GM surrounded by QF. From this expression, it is evident that the perturbative potential is influenced by several factors-such as the deformation parameter $\alpha$, the control parameter $\beta$, the symmetry breaking energy scale $\eta$, the QF parameters $(c, w)$, and the radius of curvature $\ell_p$. 

\begin{figure}[ht!]
    \centering
    \subfloat[$\alpha=5,\beta=2,c=0.01$]{\centering{}\includegraphics[width=0.45\linewidth]{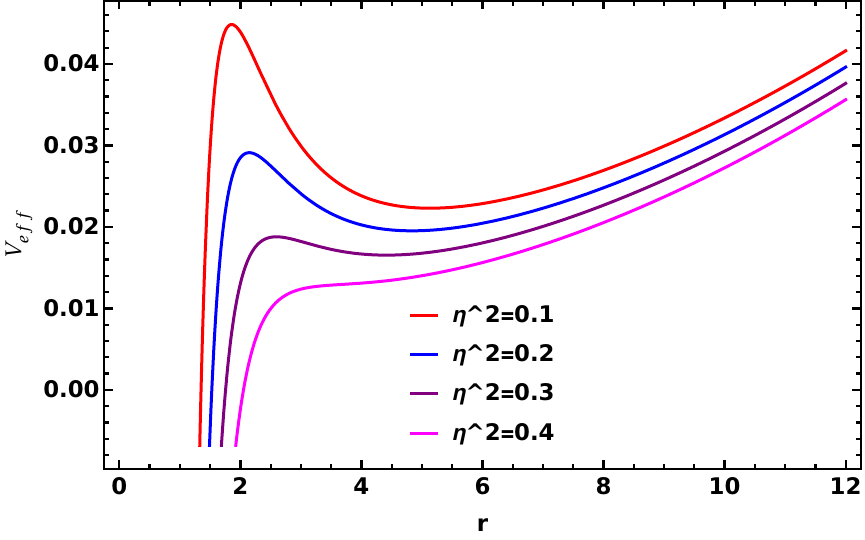}}\quad\quad
    \subfloat[$\alpha=5,\beta=2,\eta^2=0.1$]{\centering{}\includegraphics[width=0.45\linewidth]{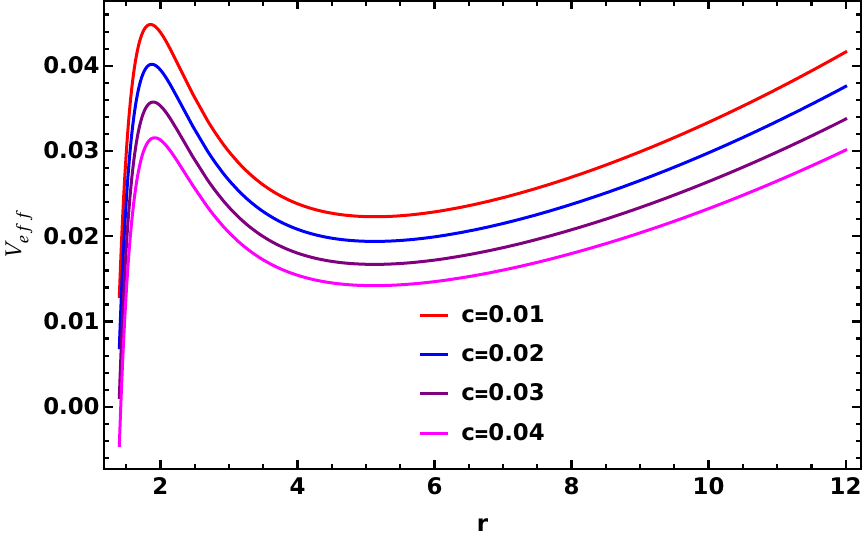}}\\
    \subfloat[$\eta^2=0.1,c=0.01$]{\centering{}\includegraphics[width=0.45\linewidth]{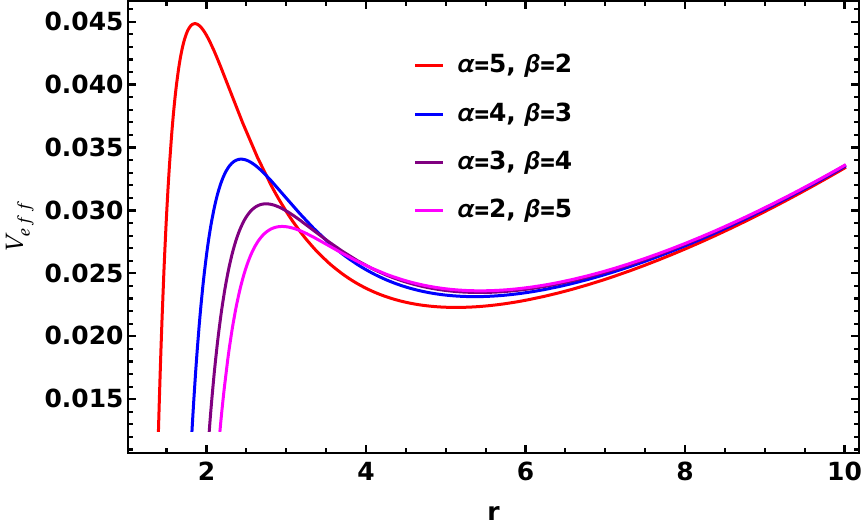}}\quad\quad
    \subfloat[$\alpha=5,\beta=2$]{\centering{}\includegraphics[width=0.45\linewidth]{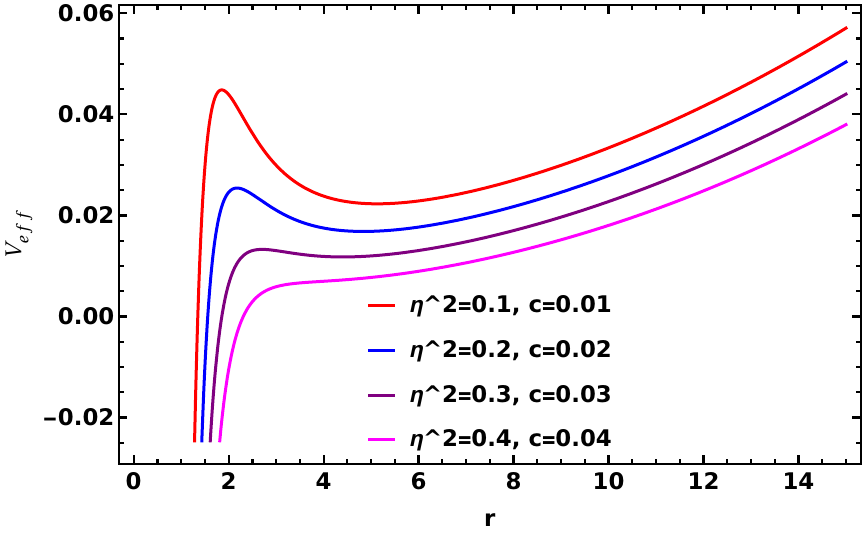}}
    \caption{The behavior of the scalar perturbative potential for different values of $\alpha, \beta, \eta$ and $c$. Here, we set $M=1$, $\ell=0$, $\Lambda=-0.03$ and $w=-2/3$.}
    \label{fig:5}
\end{figure}

For a specific state parameter, $w=-2/3$, the scalar perturbative potential (\ref{ff6}) reduces as
\begin{eqnarray}
\mathcal{V}(r)&=&\left(1-8\,\pi\,\eta^2-\frac{2\,M}{r}-c\,r+\frac{r^2}{\ell^2_{p}}+\frac{\alpha\,\beta^2}{3\,r\,(\beta+r)^3}+\frac{\alpha}{(\beta+r)^2}\right)\times\nonumber\\
&&\left\{\frac{\ell\,(\ell+1)}{r^2}+\frac{2\,M}{r^3}+\frac{2}{\ell^2_{p}}-\frac{c}{r}-\frac{\alpha\,\beta^2}{3}\,\left(\frac{1}{r^3\,(\beta+r)^3}+\frac{3}{r^2\,(\beta+r)^4}\right)-\frac{2\,\alpha}{r\,(\beta+r)^3}\right\}.\label{ff6aa}
\end{eqnarray}
In the limit $\alpha=0$, corresponding to the absence of deformation effects, the scalar perturbative potential (\ref{ff6aa}) reduces as
\begin{eqnarray}
\mathcal{V}(r)=\left(1-8\,\pi\,\eta^2-\frac{2\,M}{r}-c\,r+\frac{r^2}{\ell^2_{p}}\right)\,\left(\frac{\ell\,(\ell+1)}{r^2}+\frac{2\,M}{r^3}+\frac{2}{\ell^2_{p}}-\frac{c}{r}\right)\label{ff7}
\end{eqnarray}
Moreover, in the limit $\beta=0$, corresponding to the absence of the control parameter, the scalar perturbative potential Eq. (\ref{ff6aa}) reduces as
\begin{eqnarray}
\mathcal{V}(r)&=&\left(1-8\,\pi\,\eta^2-\frac{2\,M}{r}-c\,r+\frac{r^2}{\ell^2_{p}}+\frac{\alpha}{r^2}\right)\left\{\frac{\ell\,(\ell+1)}{r^2}+\frac{2\,M}{r^3}+\frac{2}{\ell^2_{p}}-\frac{c}{r}-\frac{2\,\alpha}{r^4}\right\}.\label{ff8}
\end{eqnarray}

In Fig.~\ref{fig:5}, we present a series of plots illustrating the behavior of the scalar perturbative potential under variations of key physical parameters. Each panel isolates the effect of a specific parameter or a combination thereof, providing insight into how these factors influence the potential.

\begin{itemize}
    \item Panel (a) shows that increasing the deformation parameter $ \alpha $ results in a noticeable reduction of the scalar potential. 
    \item Panel (b) demonstrates a similar trend, where the potential decreases as the normalization parameter $ c $ of the QF increases. 
    \item Panel (c) examines the interplay between the deformation parameter $ \alpha $ and the control parameter $ \beta $, indicating that a decrease in $ \alpha $ accompanied by an increase in $ \beta $ also leads to a lowered scalar potential.
    \item Panel (d) demonstrates that a simultaneous increase in both the symmetry-breaking energy scale $ \eta $ and the normalization parameter $ c $ results in a further suppression of the potential.     
\end{itemize} 

These results suggest that various combinations of model parameters significantly influence the scalar potential. Exploring additional parameter combinations may yield further insights, and corresponding plots can be generated to visualize their effects. Overall, the observed modifications to the potential have direct implications for the dynamics of scalar field propagation in this spacetime geometry.

\section{Electromagnetic or Vectorial Perturbations} \label{isec04}

Electromagnetic (EM) perturbations describe the behavior of test EM fields in the fixed background geometry of a BH. They offer valuable insights into BH stability, BH spectroscopy, and the interaction of BHs with external fields. These perturbations have a wide range of applications, including the study of accretion processes, jet formation, and plasma dynamics near BHs. Additionally, EM perturbations serve as useful analogs for gravitational wave behavior and play a key role in analyzing quasinormal modes (QNMs), which characterize the oscillatory response of BHs to disturbances.

For EM perturbations, the dynamics are governed by Maxwell's equations in curved spacetime:\cite{XZ,XZ2}
\begin{equation}
\frac{1}{\sqrt{-g}}\left[F_{\alpha \beta }\,g^{\alpha \nu}\,g^{\beta \mu }\sqrt{-g}\Psi\right]_{,\mu}=0,  \label{em1}
\end{equation}
where $F_{\alpha \beta }=\partial _{\alpha }A_{\beta }-\partial _{\beta}A_{\nu}$ is the EM tensor.

Following the approach adopted in \cite{XZ,XZ2}, we find
\begin{equation}
    \frac{\partial^2 \psi_\text{em}(r_*)}{\partial r^2_{*}}+\left(\omega^2-\mathcal{V}_\text{em}\right)\,\psi_\text{em}(r_*)=0,\label{em2}
\end{equation}
where the EM perturbative potential is given by
\begin{eqnarray}
\mathcal{V}_\text{em}(r)=\frac{\ell\,(\ell+1)}{r^2}\,\mathcal{F}(r)=\left(1-8\,\pi\,\eta^2-\frac{2\,M}{r}-\frac{c}{r^{3\,w+1}}+\frac{r^2}{\ell^2_{p}}+\frac{\alpha\,\beta^2}{3\,r\,(\beta+r)^3}+\frac{\alpha}{(\beta+r)^2}\right)\,\frac{\ell\,(\ell+1)}{r^2},\quad \ell\geq 1.\label{em3}
\end{eqnarray}
From the above expression (\ref{em3}), it is evident that the vectorial perturbative potential is influenced by several factors. These include the deformation parameter $\alpha$, the control parameter $\beta$, the symmetry breaking energy scale $\eta$, the QF parameters $(c, w)$, the BH mass $M$, and the radius of curvature $\ell_p$.

For a specific state parameter, $w=-2/3$, the vectorial perturbative potential (\ref{em3}) reduces as
\begin{eqnarray}
\mathcal{V}_\text{em}(r)=\left(1-8\,\pi\,\eta^2-\frac{2\,M}{r}-c\,r+\frac{r^2}{\ell^2_{p}}+\frac{\alpha\,\beta^2}{3\,r\,(\beta+r)^3}+\frac{\alpha}{(\beta+r)^2}\right)\,\frac{\ell\,(\ell+1)}{r^2}.\label{em4}
\end{eqnarray}

\begin{figure}[ht!]
    \centering
    \subfloat[$\alpha=5,\beta=2,c=0.01$]{\centering{}\includegraphics[width=0.45\linewidth]{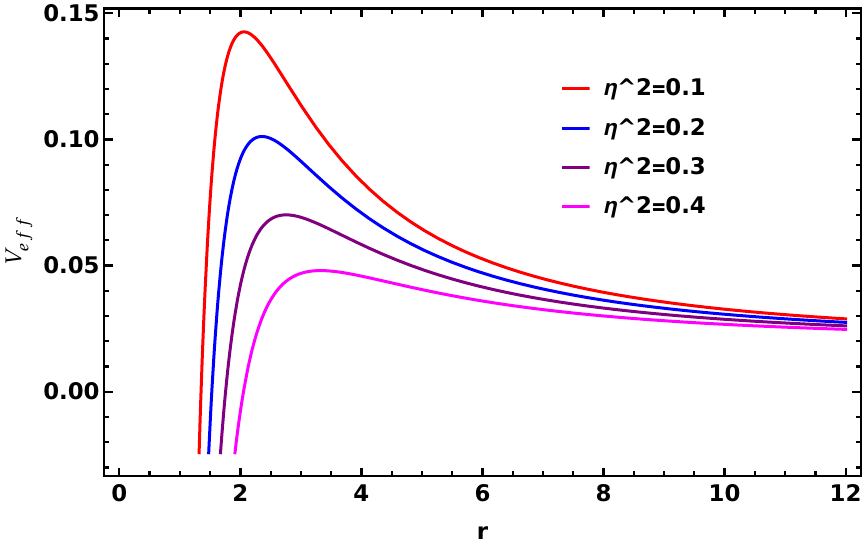}}\quad\quad
    \subfloat[$\alpha=5,\beta=2,\eta^2=0.1$]{\centering{}\includegraphics[width=0.45\linewidth]{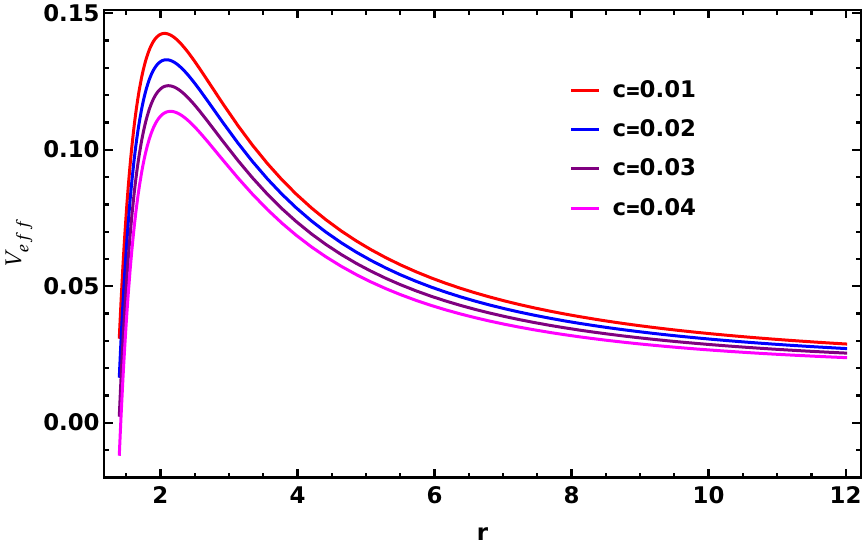}}\\
    \subfloat[$\eta^2=0.1,c=0.01$]{\centering{}\includegraphics[width=0.45\linewidth]{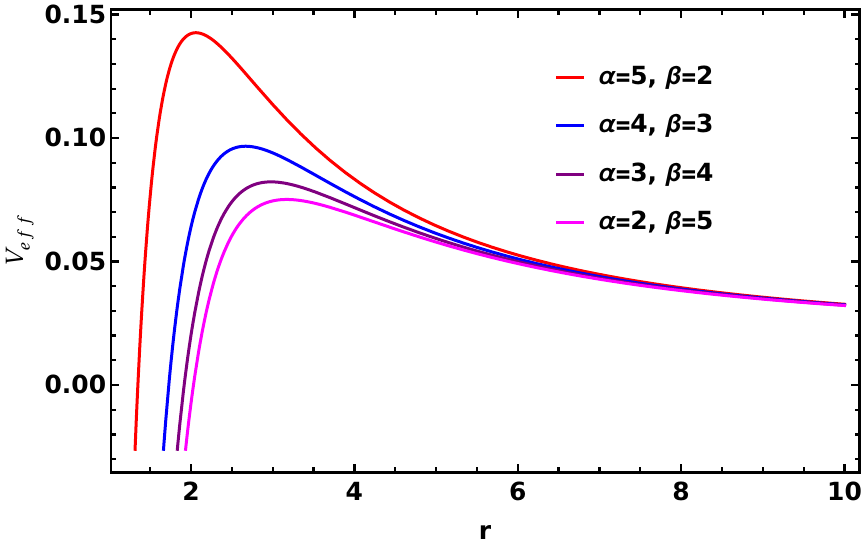}}\quad\quad
    \subfloat[$\alpha=5,\beta=2$]{\centering{}\includegraphics[width=0.45\linewidth]{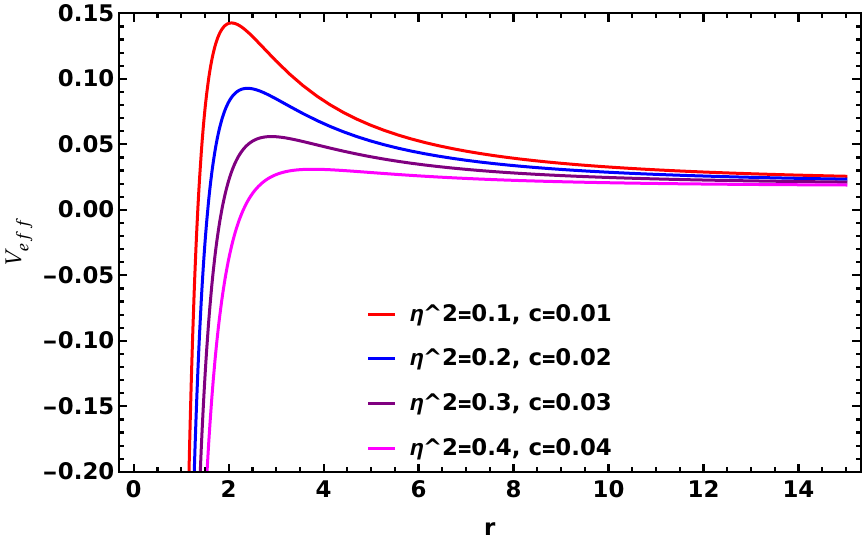}}
    \caption{The behavior of the vectorial perturbative potential for different values of $\alpha, \beta, \eta$ and $c$. Here, we set $M=1$, $\ell=1$, $\Lambda=-0.03$ and $w=-2/3$.}
    \label{fig:6}
\end{figure}

In Fig.~\ref{fig:6}, we present a series of plots illustrating the behavior of the vectorial perturbative potential under variations of key physical parameters. Each panel highlights the influence of a specific parameter or a combination of parameters, offering insights into how these factors shape the structure of the potential. Across all panels, a consistent trend is observed: the vectorial potential decreases with increasing values of certain model parameters. Specifically:
\begin{itemize}
        \item Panel (a) demonstrates that the potential decreases as the symmetry-breaking energy scale $ \eta $ increases.
        \item Panel (b) shows a similar decline in the potential with increasing values of the QF normalization constant $ c $.
        \item Panel (c) shows that a simultaneous increase in both the deformation parameter $ \alpha $ and the control parameter $ \beta $ further reduces the potential.
        \item Panel (d) indicates that jointly increasing the symmetry-breaking scale $ \eta $ and the QF normalization constant $ c $ results in an even more pronounced suppression of the potential.
\end{itemize}
    
These results underscore the sensitivity of the vectorial potential to variations in the underlying spacetime and field parameters. Investigating additional parameter combinations could provide deeper understanding, and corresponding plots can be generated to further explore these dependencies. Overall, the modifications to the vectorial potential observed here have direct implications for the propagation dynamics of electromagnetic fields and waves in this spacetime configuration.

\section{Evolution of Scalar and Vector Perturbation: Time Domain Profiles} \label{isec05}

This section presents an in-depth analysis of the time domain profiles of scalar and vector perturbations in BH spacetime, highlighting their dynamical evolution. The time domain approach provides a comprehensive perspective on how these perturbations propagate and decay over time, offering insights into transient behavior, quasinormal ring-down phases, and late-time tails. Unlike frequency domain methods, which focus on quasinormal modes (QNMs) and their spectral properties, time domain analysis captures the full evolution of perturbations, allowing for a more complete understanding of BH stability.  

To investigate these time-dependent behaviors, we employ the numerical time domain integration method as introduced by Gundlach et al. \cite{gundlach}. This method involves discretizing the tortoise coordinate $ r_* $ and time $ t $ using a uniform numerical grid, with spacing $ \Delta r_* $ and $ \Delta t $, respectively. The wave function, represented as $ \psi(r_*, t) $, evolves over this grid, with the corresponding effective potential $ V(r_*) $ influencing its behavior.  

The governing equation for scalar perturbations follows the standard wave equation in curved spacetime:  
\begin{equation}
\frac{\partial^2 \psi}{\partial t^2} - \frac{\partial^2 \psi}{\partial r_*^2} + V(r_*) \psi = 0.
\end{equation}
This equation is discretized using a second-order finite difference scheme, yielding the iterative relation:  
\begin{equation}
\psi_{i,j+1} = -\psi_{i,j-1} + \left( \frac{\Delta t}{\Delta r_*} \right)^2 (\psi_{i+1,j} + \psi_{i-1,j}) + \left[ 2 - 2\left( \frac{\Delta t}{\Delta r_*} \right)^2 - V_i \Delta t^2 \right] \psi_{i,j},
\end{equation}
where $ V_i $ is the effective potential evaluated at grid point $ i $. This formulation allows us to iteratively determine the wave function’s evolution at successive time steps.  

For the initial conditions, we assume a Gaussian wave packet centered at $ r_* = k_1 $ with width $ \sigma $:  
\begin{equation}
\psi(r_*, t=0) = \exp \left[ -\frac{(r_* - k_1)^2}{2\sigma^2} \right].
\end{equation}
For $ t < 0 $, the perturbation is set to zero. This localized initial disturbance provides a clear means to track the perturbation’s propagation and interaction with the BH's effective potential.  

Numerical stability is ensured through the Von Neumann stability criterion, which requires the ratio $ \frac{\Delta t}{\Delta r_*} $ to be less than 1. In our implementation, we set $ \frac{\Delta t}{\Delta r_*} = 0.5 $, ensuring both numerical stability and computational efficiency. The computational domain extends sufficiently far in $ r_* $ to encompass both near-horizon and asymptotic regions, allowing for an accurate depiction of perturbation evolution.  

The generated time domain profiles illustrate the distinct evolutionary stages of the perturbation. Initially, a transient phase occurs as the wave packet adapts to the BH’s spacetime structure. This is followed by the quasinormal ring-down phase, characterized by damped oscillations corresponding to the BH’s QNMs.

\begin{figure}[htbp]
\centerline{
   \includegraphics[scale = 0.68]{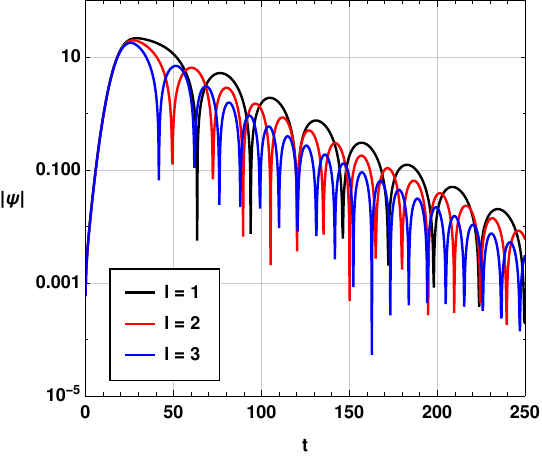}\hspace{0.5cm}
   \includegraphics[scale = 0.68]{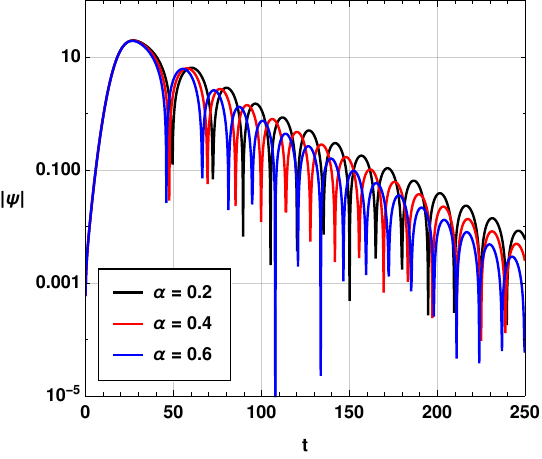}} \vspace{-0.2cm}
\caption{The time-domain profiles of the massless scalar
perturbations with the parameter values $M=1$, $G = 1$, $8 \pi = 1$, $n= 0$, $c = 0.1$, $\beta = 0.3$, $\eta = 0.1$ and $\Lambda = -0.002$. On the first panel, $\alpha =0.2$, and on the second panel, $l =2$. }
\label{time01}
\end{figure}

\begin{figure}[htbp]
\centerline{
   \includegraphics[scale = 0.68]{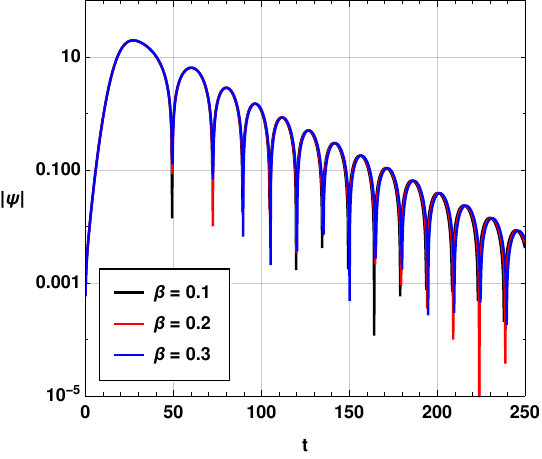}\hspace{0.5cm}
   \includegraphics[scale = 0.68]{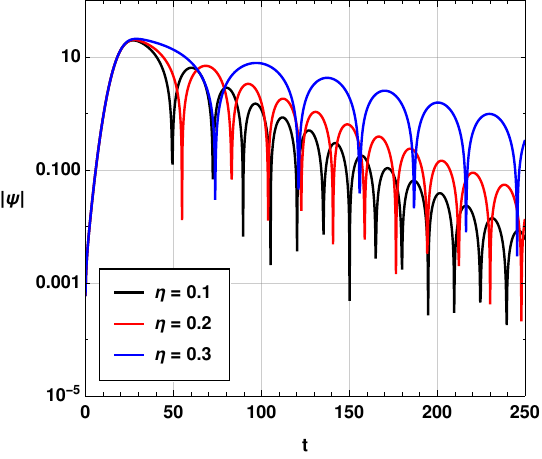}} 
   \includegraphics[scale=0.68]{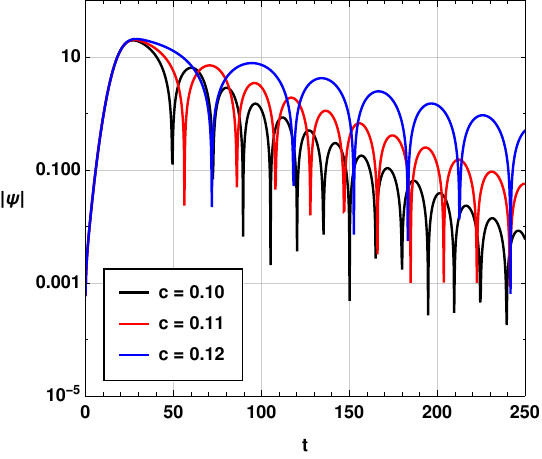}\vspace{-0.2cm}
\caption{The time-domain profiles of the massless scalar
perturbations with the parameter values $M=1$, $G = 1$, $8 \pi = 1$, $n= 0$, $l = 2$, $\alpha = 0.2$, and $\Lambda = -0.002$. On the first panel, $\eta =0.1$, $c=0.1$, on the second panel, $\beta =0.3$, $c=0.1$ and on the third panel, $\eta =0.1$, $\beta=0.3$. }
\label{time02}
\end{figure}

\begin{figure}[htbp]
\centerline{
   \includegraphics[scale = 0.7]{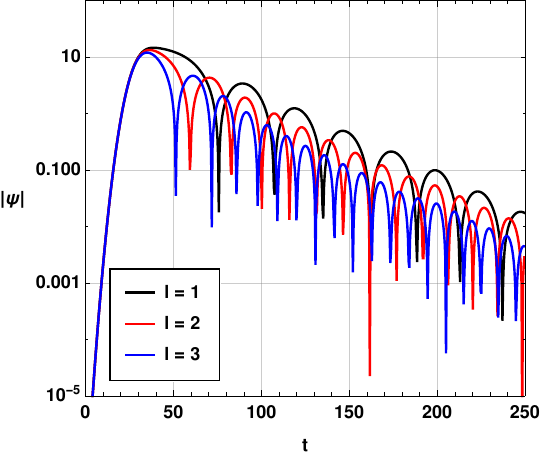}\hspace{0.5cm}
   \includegraphics[scale = 0.7]{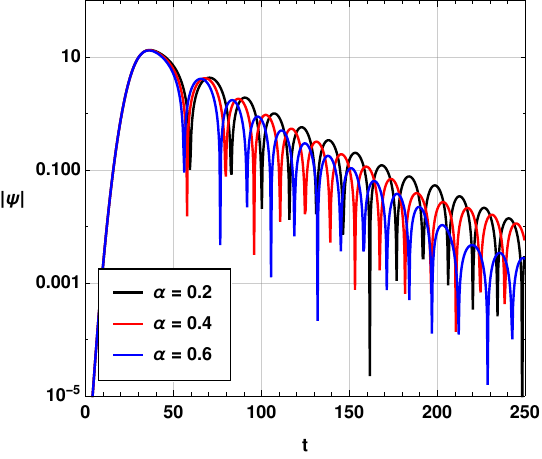}} \vspace{-0.2cm}
\caption{The time-domain profiles of the vector
perturbations with the parameter values $M=1$, $G = 1$, $8 \pi = 1$, $n= 0$, $c = 0.1$, $\beta = 0.3$, $\eta = 0.1$ and $\Lambda = -0.002$. Left panel, $\alpha =0.2$, and right panel, $\ell =2$. }
\label{time03}
\end{figure}

\begin{figure}[htbp]
    \centerline{
   \includegraphics[scale = 0.7]{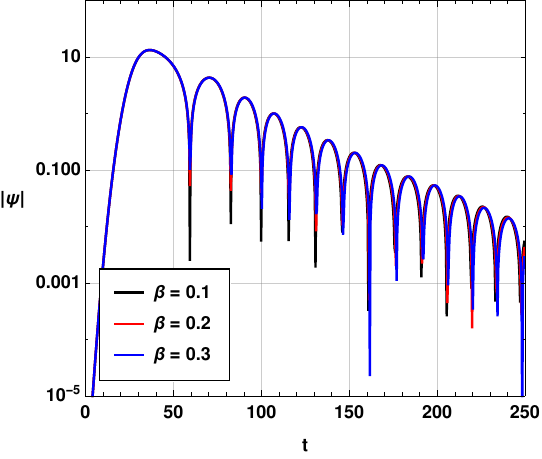}\hspace{0.5cm}\quad
   \includegraphics[scale = 0.7]{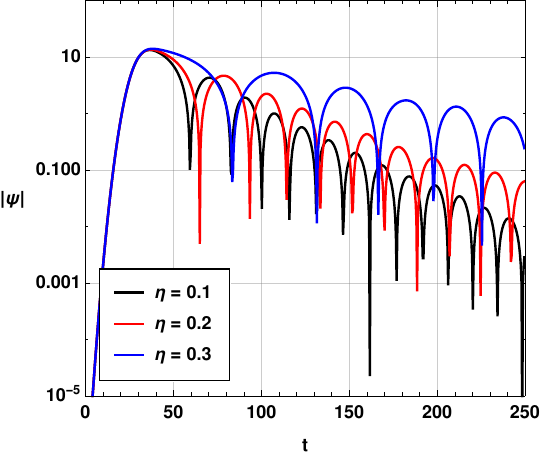}}
   \includegraphics[scale=0.7]{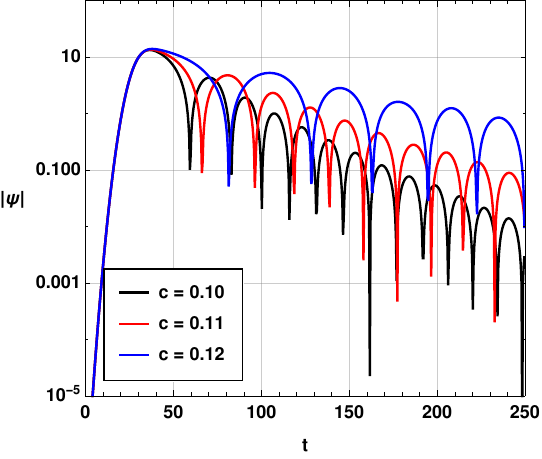}\vspace{-0.2cm}
\caption{The time-domain profiles of the vector
perturbations with the parameter values $M=1$, $G = 1$, $8 \pi = 1$, $n= 0$, $l = 2$, $\alpha = 0.2$, and $\Lambda = -0.002$. On the first panel, $\eta =0.1$, $c=0.1$, on the second panel, $\beta =0.3$, $c=0.1$ and on the third panel, $\eta =0.1$, $\beta=0.3$. }
\label{time04}
\end{figure}

In Fig.s \ref{time01} and \ref{time02}, we show the time domain profiles for scalar perturbations for different sets of model parameters. From the left panel of Fig. \ref{time01}, one may see that the multipole moment $l$ has a noticeable impact on the time domain profile of the ring-down gravitational waves (GWs). With an increase in the value of $l$, we observe an increase in the oscillation frequency and damping rate as well. On the left panel of Fig. \ref{time01}, we have shown the impacts of model parameter $\alpha$ on the time domain profiles. The parameter $\alpha$ significantly impacts both damping rate and the oscillation frequency of the ring-down GWs. On the first panel of Fig. \ref{time02}, we have shown the variation of time domain profiles with respect to the parameter $\beta$ and find that the impact of the parameter $\beta$ is negligible on the ring-down phase of GWs. However, from the second and third panel of Fig. \ref{time02}, we see that the parameters $\eta$ and $c$ impact the evolution of scalar perturbations significantly and in both the cases, an increase in the value of the parameter results in a decrease in the damping rate and oscillation frequency of ring-down GWs.

In Fig. \ref{time03} and \ref{time04}, the time domain profiles of electromagnetic or vector perturbations are shown. It can be seen that the behavior of time domain profiles of vector perturbations are similar with a slight variation in the oscillation frequency and damping rates. 

\section{Emission Rate} \label{isec06}

The emission rate of a BH plays a fundamental role in understanding its interaction with quantum fields and radiation processes. One of the key observables associated with this phenomenon is the BH shadow, which serves as a crucial link between the absorption cross-section and the energy emission rate. For a distant observer, the BH shadow manifests as a dark region against the surrounding radiation field, effectively representing the high-energy absorption cross-section \cite{Wei:2013kza}. This feature not only provides insights into BH properties but also influences the rate at which it emits particles through quantum effects.  

In the case of a spherically symmetric BH, such as the Schwarzschild BH, the absorption cross-section at high energies exhibits an oscillatory behavior around a limiting value, denoted as $ \sigma_{\text{lim}} $. This limiting cross-section is directly related to the BH shadow and can be approximated by:  

\begin{equation}
\sigma_{\text{lim}} \approx \pi R_s^2,
\end{equation}

where $ R_s $ is the shadow radius. The photon sphere, which forms the boundary of the shadow, dictates the effective size of the absorption cross-section. For a Schwarzschild BH, the photon sphere is located at $ r = 3M $, where $ M $ represents the BH mass in geometric units. This relation establishes a direct connection between the BH’s geometry and its interaction with external particles.  

Beyond its role in absorption, the shadow also influences the BH's emission rate, particularly in the context of Hawking radiation. This thermal radiation arises due to quantum fluctuations near the event horizon, allowing BHs to slowly lose mass over time. The energy emission rate per unit frequency is given by \cite{Wei:2013kza, EslamPanah:2020hoj, Kruglov:2021qzd, Kruglov:2021stm, Belhaj:2020rdb}:  

\begin{equation}
E_R (\omega) = \frac{d^2 \mathcal{E}(\omega)}{dt \, d\omega} = \frac{2\pi^3 \omega^3 R_s^2}{e^{\frac{\omega}{T_H}} - 1},
\end{equation}

where $ \omega $ is the frequency of the emitted radiation, and $ T_H $ is the Hawking temperature, which is determined by the surface gravity at the event horizon:  

\begin{equation}
T_H = \frac{\hbar  \mathcal{F}'(r_h)}{4 \pi}.
\end{equation}

Here, $ r_h $ denotes the event horizon radius, and $  \mathcal{F}'(r_h) $ is the derivative of the BH metric function evaluated at $ r_h $. For a Schwarzschild BH, with the metric function $ f(r) = 1 - \frac{2M}{r} $, we obtain $ f'(r_h) = \frac{1}{2M} $ at $ r_h = 2M $, leading to a well-defined expression for the Hawking temperature.  

\begin{figure}[htbp]
\centerline{
   \includegraphics[scale = 0.85]{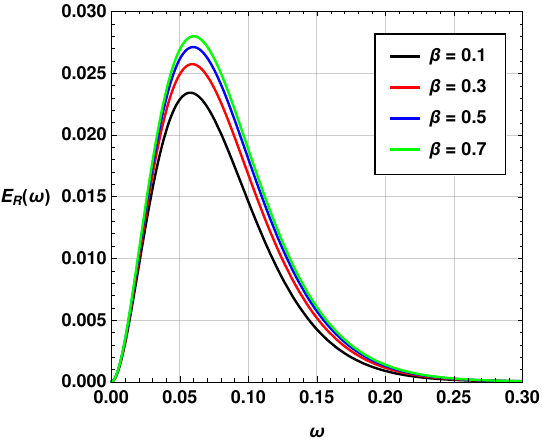}\hspace{0.5cm}
   \includegraphics[scale = 0.83]{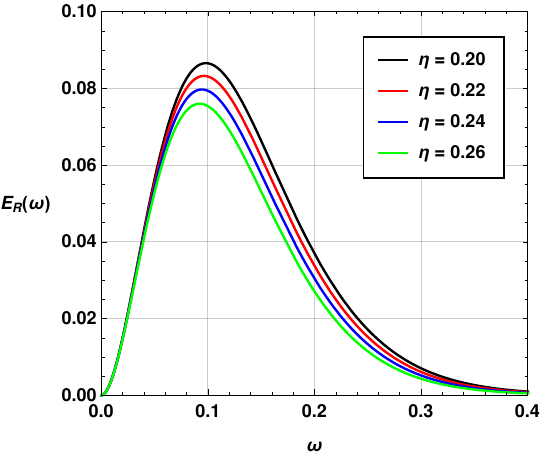}} 
   \vspace{-0.2cm}
\caption{The emission rate of the black hole with $M=8\pi=1, c=0.01$ and $\Lambda=-0.002$. On the first panel, $\alpha=0.7,  \eta=0.5$ and on the second panel $\alpha=0.1, \beta=0.1$.}
\label{EmR}
\end{figure}

\begin{figure}[htbp]
\centerline{
   \includegraphics[scale = 0.85]{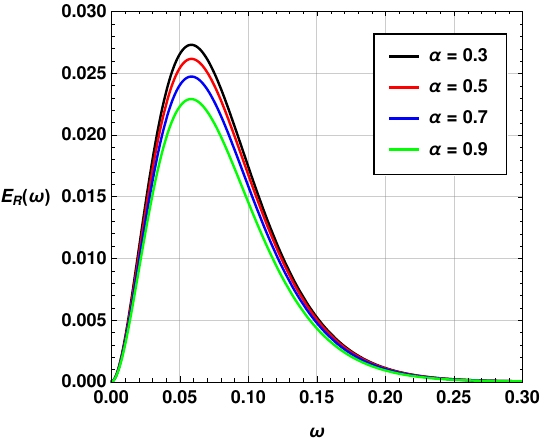}\hspace{0.5cm}
   \includegraphics[scale = 0.83]{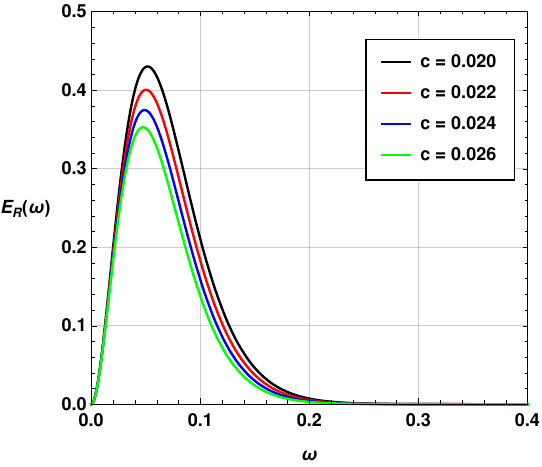}} 
   \vspace{-0.2cm}
\caption{The emission rate of the black hole with $M=8\pi=1, \eta=0.5$ and $\Lambda=-0.002$. On the first panel, $\beta=0.2, c=0.01$ and on the second panel $\alpha=0.3, \beta=0.2$.}
\label{EmR2}
\end{figure}

The dependence of the emission rate on $ R_s^2 $ highlights the significance of the BH shadow in determining the intensity of Hawking radiation. A larger shadow radius corresponds to a higher emission rate for a given frequency, reinforcing the geometric connection between the shadow and the BH’s thermal properties. Meanwhile, the spectral distribution of the emitted radiation is governed by the term $ e^{\frac{\omega}{T_H}} - 1 $, which dictates the peak emission frequency based on the Hawking temperature.  

Thus, the emission rate of a BH is intrinsically linked to its shadow and the underlying quantum processes that govern Hawking radiation. The shadow, acting as a geometric marker of the absorption cross-section, provides a crucial parameter for understanding how BHs interact with and emit radiation. This connection bridges classical BH properties with quantum field theory, making the emission rate a key quantity in both theoretical studies and astrophysical observations.  

In Fig.s \ref{EmR} and \ref{EmR2}, we have shown the variation of the emission rate of the BH for a different set of model parameters. On the left panel of Fig. \ref{EmR}, we have shown the impacts of $\beta$ on the emission rate of the BH. One may note that with an increase in the value of the parameter $\beta$, the emission rate of the BH increases, decreasing its lifetime. On the other hand, the right panel of Fig. \ref{EmR} depicts that the parameter $\eta$ has an opposite impact on the emission rate of the BH. An increase in the model parameter $\eta$ results in a reduced BH emission rate, thereby extending its lifespan. Thus, $\eta$ enhances BH stability, whereas the control parameter, linked to the absence of a central singularity, destabilizes it by boosting the emission rate.

On the first panel of Fig. \ref{EmR2}, we have shown the impact of $\alpha$ on the emission rate of the black hole. One can see that an increase in the value of the model parameter $\alpha$ results in a decrease in the emission rate. The second panel of Fig. \ref{EmR2} shows that an increase in the value of the parameter $c$ also has a similar impact on the emission rate of the black hole. These results show that the deformation parameter $\alpha$ is responsible for the increase in the lifetime of the black hole by decreasing the emission rate. The presence of the quintessence field also extends the lifespan of the black hole.

\section{Concluding Remarks}\label{isec07}

In this study, we investigated the geodesic structure and perturbative properties of a deformed AdS-Schwarzschild BH with GM surrounded by QF. Our analysis provided significant insights into how the deformation parameter $\alpha$, the control parameter $\beta$, the GM parameter $\eta$, and the QF parameter $c$ collectively influence the physical properties of this modified BH geometry.

We began by examining the geodesic motion of both null and timelike particles in this BH spacetime. For null geodesics, we determined the photon sphere radius through numerical calculations and presented the results in Tables \ref{table1a}--\ref{table3a}. Our findings revealed that the photon sphere expands with increasing GM parameter $\eta$, as clearly illustrated in Fig. \ref{figa1}, establishing the significant role of GMs in shaping the BH's optical properties. Conversely, we observed that increasing the deformation parameter $\alpha$ reduces the photon sphere radius, while increasing the control parameter $\beta$ enlarges it, as demonstrated in Fig. \ref{figa2}. These parameter-dependent variations of the photon sphere have direct implications for the BH shadow, which has become increasingly relevant in the era of direct BH imaging through projects like the Event Horizon Telescope \cite{isz15,isz16}.

Our analysis of the forces acting on photon particles, as given by Eq. (\ref{dd6}), demonstrated how the parameters $\alpha$, $\beta$, $\eta$, and $c$ alter the photon dynamics in the BH's gravitational field. This was visually represented in Fig. \ref{fig:2}, where we observed systematic changes in the force profiles with variations in these parameters. Additionally, we computed the Lyapunov exponent for circular null orbits using Eq. (\ref{lyapunov4}) and investigated its behavior across different mass regimes. The results illustrated in Fig. \ref{fig:3} revealed a remarkable transition from unstable to stable circular null geodesics as the BH mass increases, with important implications for quasinormal mode spectra. For timelike geodesics, we derived expressions for the energy and angular momentum of massive test particles in circular orbits, as given by Eq.s (\ref{ee4}) and (\ref{ee5}). The effective potential, displayed in Fig. \ref{fig:4}, exhibited significant parameter-dependent variations, indicating how the parameters $\alpha$, $\beta$, $\eta$, and $c$ influence the stability and characteristics of orbital motion for massive bodies. These results provide valuable insights into matter accretion processes around such modified BHs \cite{isz37}.

A major component of our study involved the analysis of perturbative potentials for both scalar and electromagnetic fields in this BH background. The scalar perturbative potential, given by Eq. (\ref{ff6}), showed remarkable sensitivity to the model parameters, as illustrated in Fig. \ref{fig:5}. Similarly, the electromagnetic perturbative potential, described by Eq. (\ref{em3}), exhibited systematic variations with changes in $\alpha$, $\beta$, $\eta$, and $c$, as shown in Fig. \ref{fig:6}. These results are particularly significant for understanding the stability of the BH under various types of perturbations and for characterizing its quasinormal mode spectrum \cite{isz17,isz18}. Our time-domain analysis of scalar and vector perturbations provided further insights into the dynamical evolution of disturbances in this BH spacetime. Using the numerical integration method developed by Gundlach et al. \cite{isz25}, we generated time-domain profiles that clearly illustrated the distinct evolutionary stages: the initial transient phase, the quasinormal ringing phase dominated by the BH's characteristic frequencies, and the late-time power-law decay phase. These profiles, shown in Fig.s \ref{time01}, \ref{time02}, \ref{time03} and \ref{time04}, demonstrated how the model parameters influence the damping rates and oscillation frequencies, thus affecting the BH's response to external perturbations.

Finally, we investigated the emission rate of the BH, establishing connections between its shadow radius and thermal properties. The results presented in Fig.s \ref{EmR} and \ref{EmR2} revealed that the control parameter $\beta$ increases the emission rate, thereby reducing the BH's lifetime, while the GM parameter $\eta$, deformation parameter $\alpha$ and QF parameter $c$ have the opposite effect, enhancing BH stability by suppressing emission. These findings provide valuable insights into the thermodynamic behavior of deformed BHs with GMs and QFs, potentially offering glimpses into quantum gravity effects \cite{isz36}. Throughout our analysis, we consistently observed that the deformation parameter $\alpha$ and the control parameter $\beta$, which arise from theoretical considerations related to quantum gravity and the avoidance of central singularities, significantly modify the classical properties of the Schwarzschild BH. Similarly, the GM parameter $\eta$ and the QF parameter $c$, which have cosmological origins, introduce substantial changes to the BH geometry and its associated physical properties. The results of our study contribute to the broader understanding of modified BH solutions and their implications for fundamental physics. The parameter-dependent variations in geodesic structure, perturbative properties, and thermodynamic characteristics offer potential observational signatures that could be tested against astrophysical data, thereby providing constraints on these theoretical models \cite{isz38,isz39}. Moreover, our findings on the stability and emission properties of these BHs have relevance for cosmic censorship considerations and BH evaporation scenarios \cite{isz40}.

Looking ahead, several promising avenues for future research emerge from our current work. First, extending our analysis to rotating BH solutions would incorporate the effects of spin, which is crucial for realistic astrophysical BHs. Second, investigating the implications of our findings for gravitational wave signals from BH mergers involving such modified BHs could provide observational tests for these theoretical models. Third, further exploration of the thermodynamic phase structure in the extended phase space, incorporating the effects of thermal fluctuations more comprehensively, could yield deeper insights into quantum aspects of BH physics. Finally, applying our methodologies to other modified gravity theories and alternative BH solutions could help establish a more complete understanding of how quantum gravity effects manifest in BH physics.

\section*{Acknowledgments}
 F.A. gratefully acknowledges the Inter University Centre for Astronomy and Astrophysics (IUCAA), Pune, India, for the conferment of a visiting associateship.  \.{I}.~S. extends sincere thanks to TÜBİTAK, ANKOS, and SCOAP3 for their support in facilitating networking activities under COST Actions CA22113, CA21106, and CA23130.

\section*{Data Availability Statement}
There are no new data associated with this article.

\end{document}